\documentclass[a4paper,11pt]{article}
\pdfoutput=1 

\usepackage[T1]{fontenc} 
\usepackage{lineno}
\usepackage{url}

\usepackage{blindtext}
\usepackage{amsmath}
\usepackage{amssymb}
\usepackage{epsfig}
\usepackage{graphicx}
\usepackage{amsmath}
\usepackage{geometry}
\usepackage{graphicx}
\usepackage{microtype}
\usepackage{multirow}
\usepackage{float}
\usepackage{rotating}
\usepackage[section]{placeins}
\usepackage{anysize}
\usepackage{enumerate}
\usepackage{subcaption}

\usepackage{color}

\usepackage{xspace}

\providecommand{\PYTHIA} {{\textsc{pythia}}\xspace}
\providecommand{\SHERPA} {{\textsc{sherpa}}\xspace}

\providecommand{\PROFESSOR} {\textsc{professor}\xspace}
\providecommand{\RIVET} {\textsc{RIVET}\xspace}

\newcommand{\alpS}{\ensuremath{\alpha_{s}}\xspace}

\newcommand{\rpv}{\ensuremath{\rlap{\kern.2em/}R}\xspace}

\usepackage{url}
\usepackage{soul}

\DeclareFixedFont\trfont{OT1}{phv}{b}{sc}{11}
\newif\ifnotoc\notocfalse
\newif\ifemailadd\emailaddfalse
\newif\iftoccontinuous\toccontinuousfalse
\newif\ifnatbibsort\natbibsorttrue

\relax

\RequirePackage{amsmath}
\RequirePackage{amssymb}
\RequirePackage{epsfig}
\RequirePackage{graphicx}
\ifnatbibsort\RequirePackage[numbers,sort&compress]{natbib}\else\RequirePackage[numbers,compress]{natbib}\fi
\RequirePackage[colorlinks=true
  ,urlcolor=blue
  ,anchorcolor=blue
  ,citecolor=blue
  ,filecolor=blue
  ,linkcolor=blue
  ,menucolor=blue
  ,pagecolor=blue
  ,linktocpage=true
  ,pdfproducer=medialab
  ,pdfa=true
]{hyperref}

\renewcommand{\theequation}{\thesection.\arabic{equation}}
\numberwithin{equation}{section}
\def\fnum@figure{\textbf{\figurename\nobreakspace\thefigure}}
\def\fnum@table{\textbf{\tablename\nobreakspace\thetable}}

\renewenvironment{thebibliography}[1]{%
\begin{oldthebibliography}{#1}%
\small%
\raggedright%
\setlength{\itemsep}{5pt plus 0.2ex minus 0.05ex}%
}%
{%
\end{oldthebibliography}%
}

\usepackage{xspace}
\title{\bf Optimizing the parton shower model in PYTHIA with pp collision data at \boldmath{ $\mathrm{\surd{s}=13\, TeV}$}}
\author{S. K. Kundu\thanks{skundu91phys@gmail.com},    T. Sarkar\thanks{tanmayvb@gmail.com},   M. Maity\thanks{manas.maity@visva-bharati.ac.in}  }

\date{%
    Department of Physics, Visva-Bharati University \\Santiniketan, India\\%
}

\newcommand{\PY}        {\textsc{pythia8}\xspace}
\newcommand{\HWpp}      {\textsc{herwig++}\xspace}
\newcommand{\MGNLO}     {\textsc{MadGraph5}\_a\textsc{mc@nlo}\xspace }
\newcommand{\cindf}     {\ensuremath{\chi^{2}/NDF}\xspace}  

\newcommand{\pT}        {p_\textrm{T} }
\newcommand{\taup}      {\tau_{\perp}}
\newcommand{\mass}      {\rho_{\text{Tot}}}
\newcommand{\massp}     {\rho^{\textrm{T}}_{\text{Tot}}}
\newcommand{\bT}        {B_\textrm{T}}
\newcommand{\HT}        {H_{\textrm{T,2}}}
\newcommand{\asISR}     {\alpha_{s}^{\textrm{ISR}}} 
\newcommand{\asFSR}     {\alpha_{s}^{\textrm{FSR}}} 
\newcommand{\ak}        {\textrm{anti-}k_{\textrm{T}}}
\newcommand{\Pevol}     {p_{\perp}}

\begin{document} 
\maketitle

\begin{abstract}
Production of quarks and gluons in hadron collisions tests Quantum 
Chromodynamics (QCD) over a wide range of energy. Models of QCD are implemented 
in event generators to simulate hadron collisions and evolution of quarks and gluons 
into jets of hadrons. \PY uses the parton shower model for simulating particle 
collisions and is optimized using experimental observations. Recent measurements of 
event shape variables and jet cross-sections in pp collisions at $\mathrm{\surd{s} = 13\, TeV}$ 
at the Large Hadron Collider have been used to optimize the parton shower model as 
used in \PY.
\end{abstract}



\section{Introduction}\label{sec:1}
High energy collisions of hadrons produce quarks and gluons, collectively known as
partons, which evolve into a large number of stable hadrons. Successive stages of this
process involve lower energy and hence increasing value of the strong coupling
$\mathrm{\alpS}$. This makes analytical calculations based on perturbation theory
prohibitively difficult as well as unreliable. Thus, a combination of analytical calculations
in the early stages, including the hard scattering, and approximate numerical calculations
later, is used to describe such events.

A number of event shape variables (ESVs) \cite{Banfi:2004nk,Banfi:2010xy} have been defined to reflect the different
aspects of QCD over a wide range of energy as well as theoretical complexity. Their measurements\cite{Sirunyan:2018adt,
Aad:2016ria, Khachatryan:2014ika,Chatrchyan:2013tna, Aad:2012np, Aad:2012fza, Abelev:2012sk, Khachatryan:2011dx} at
the colliders have been used to improve our understanding of QCD - parton distribution function
(PDF), multi-parton interaction (MPI), initial state radiation(ISR), final state radiation (FSR),
fragmentation, hadronization, etc. Although data from LEP, Tevatron and RunI of the Large Hadron
Collider(LHC) have been used to this end, RunII data of LHC has so far not been used exhaustively.
This study attempts to improve the understanding of QCD using four ESVs measured in RunII data
\cite{Sirunyan:2018adt}.

Monte Carlo (MC) simulation programs such as \PY~\cite{Sjostrand:2014zea},
\HWpp~\cite{Bellm:2015jjp}, \MGNLO~\cite{Alwall:2011uj}, etc. combine analytical calculations
and approximate models to describe the collisions. \PY is a popular MC event generator and is
extensively used by the particle physics community. Event generators which use matrix
element (ME) calculations for the hard scattering process often use \PY for emulating the
subsequent fragmentation and hadronization process. Data from colliders have been used to optimize
\PYTHIA ~\cite{Skands:2014pea, Abreu:1996na, Gunnellini:2018kug}and it is imperative that data from the RunII of LHC is used to optimize it.

The plan of the paper is as follows. Section \ref{sec:2} briefly reviews the theory
relevant to parton shower. Section \ref{sec:3} examines the agreement of \PY with
measurement of ESVs by the CMS experiment \cite{Sirunyan:2018adt} in
pp collisions at $\mathrm{\surd{s} = 13}$ TeV. Section \ref{sec:4} describes the studies
of the different parameters of \PY and optimization of the parameter set. It also describes
the application of the improved parameter set to compare with results of other analyses of
CMS \cite{Khachatryan:2016wdh} and ATLAS \cite{ATLAS:2016djc}. Section \ref{sec:5}
contains the summary and outlook.

\section{Parton Shower description of hadron collisions}\label{sec:2}
In high energy collisions, it is intensely complicated to achieve an exclusive picture of the final
state partons using matrix elements calculation and only fixed order treatment is not sufficient. 
However, for comparison with experimental analyses, a fully exclusive description of the final states using 
calculations based on the shower evolution and hadronization is more suitable. Such methods are described 
through phenomenological models embedded in the shower MC codes. MC event generators such as \PY, \HWpp and 
\SHERPA use their own parton shower treatment which have their own merits. \HWpp uses the coherent branching 
algorithm \cite{Banfi:2006gy} for angular order branching whereas Catani-Seymour dipole factorisation 
\cite{Schumann:2007mg} is used in \SHERPA.

\PY uses leading order(LO) calculations for generating the $2\rightarrow 2$ hard scattering 
processes. It uses `transverse momentum' ordered parton shower\cite{Sjostrand:2004ef} 
with  $p^{2}_{\perp}$ as evolution variable for the generation of $2\rightarrow n$ 
($n\ge 2$) final states by taking account ISR and FSR. \PY also emulates MPI and 
evolution of the partons into hadrons. In this showering scheme, resummation is done to 
all orders with a certain logarithmic accuracy.

For the splitting of a parton $a\rightarrow bc$, \PY uses the branching probability expressed 
by the DGLAP evolution equations:
\begin{equation}
d\mathcal{P}_{a}=\frac{dp^{2}_{\perp}}{p^{2}_{\perp}}\sum_{b,c}\frac{\alpha_{s}(p^{2}_{\perp})}{2\pi}P_{a\rightarrow bc}(z)dz 
\label{DGLAP}
\end{equation}
where $\mathrm{P_{a\rightarrow bc}}$ is the DGLAP splitting function and $p^{2}_{\perp}$ represents the 
scale of the branching; $z$ represents the sharing of $\Pevol$ of $a$ between the 
two daughters, with $b$ taking a fraction $z$ and $c$ the rest, $1-z$. Here the summation goes over 
all allowed branching, e.g. $q\rightarrow qg$ and $\mathrm{q\rightarrow q\gamma}$ and so on. Now, 
these probabilities become larger than unity due to divergence when $p^{2}_{\perp}\rightarrow 0$ 
which is taken care by introducing a term $\mathcal{P}^{no}_a(p^{2}_{{\perp}_{\textrm{max}}},p^{2}_{{\perp}_{\textrm{evol}}})$
known as \emph{Sudakov form factor} \cite{Sudakov:1954sw}. This Sudakov factor ensures that there will be no emission between scale
$p^2_{{\perp}_{\textrm{max}}}$ to a given $p^2_{{\perp}_{\textrm{evol}}}$.

Considering lightcone kinematics, evolution variables for $a\rightarrow bc$ at virtuality scale $Q^2$ 
for space-like branching (ISR) and time-like branching (FSR) are given by equations \ref{ISR_PT} and 
\ref{FSR_PT}, respectively.
\begin{eqnarray}
p^{2}_{{\perp}_{\textrm{evol}}} & = & (1-z)Q^2 \label{ISR_PT} \\
p^{2}_{{\perp}_{\textrm{evol}}} & = & z(1-z)Q^2 \label{FSR_PT} 
\end{eqnarray}
Finally, equations \ref{ISR_evol} and \ref{FSR_evol} describe the evolutions for ISR and FSR respectively \cite{Sjostrand:2004ef}.
\begin{eqnarray}
\textrm{d}\mathcal{P}_{b} & = & \frac{\textrm{d}p^{2}_{{\perp}_{\textrm{evol}}}}{p^{2}_{{\perp}_{\textrm{evol}}}}\frac{\alpha_s(p^{2}_{{\perp}_{\textrm{evol}}})}{2\pi}
	\frac{x^\prime f_a (x^\prime , p^{2}_{{\perp}_{\textrm{evol}}})}{xf_a (x, p^{2}_{{\perp}_{\textrm{evol}}})}
	P_{a\rightarrow bc}(z) dz \mathcal{P}^{\textrm{no}}_b (x,p^{2}_{{\perp}_{\textrm{max}}},p^{2}_{{\perp}_{\textrm{evol}}}) \label{ISR_evol} \\
\textrm{d}\mathcal{P}_{a} & = & \frac{\textrm{d}p^{2}_{{\perp}_{\textrm{evol}}}}{p^{2}_{{\perp}_{\textrm{evol}}}}\frac{\alpha_s(p^{2}_{{\perp }_{\textrm{evol}}})}{2\pi}P_{a\rightarrow bc}(z) \textrm{d}z \mathcal{P}^{\textrm{no}}_a (p^{2}_{{\perp}_{\textrm{max}}},p^{2}_{{\perp}_{\textrm{evol}}})  \label{FSR_evol}
\end{eqnarray}

 Currently both the running re-normalisation and factorisation shower scales, i.e. the scales at 
 which  $\alpS$ and the PDFs are evaluated, are chosen to be $p^2_{\perp_\textrm{evol}}$ \cite{Corke:2010yf}. The general 
 methodology of \PY for ISR, FSR, MPI is to start from some maximum scale $p^{2}_{{\perp}_{\textrm{max}}}$ 
 and evolve downward in energies towards next branching untill the daughter partons reach some cut-off.
 

\section{Event Shape Variables at RunII of LHC}\label{sec:3}
Event shape variables\cite{Banfi:2004nk}  are defined in terms of the four-momenta of the objects 
with multi-jet final states. They are sensitive to the 
topology of the primary hard scattering and also, the evolution of the primary partons 
into stable hadrons. With a judicious choice of the ESVs it is possible to use experimental 
data to confront QCD predictions - from perturbative analytical calculation in the early phase to the nonperturbative 
models in the later phase of the event. These variables also have the advantage that, 
they are safe from collinear and infrared(IRC) divergences \cite{Banfi:2010xy}. 
The ESVs may also be used to look for signature of physics beyond the Standard Model 
\cite{Chatterjee:2012qt, Datta:2011vg}. LHC experiments ATLAS, CMS, and ALICE have 
studied various ESVs, evaluating them with jets and charged particles in proton-proton (pp) collision
\cite{Sirunyan:2018adt,
Aad:2016ria, Khachatryan:2014ika,Chatrchyan:2013tna, Aad:2012np, Aad:2012fza, Abelev:2012sk, Khachatryan:2011dx}. 

Most recently the CMS has evaluated four ESVs in multi-jet events (with atleast three jets)
in pp collisions at $\surd{s} = 13$ TeV \cite{Sirunyan:2018adt}. These are - the complement of 
transverse thrust ($\mathrm{\taup}$), total jet mass ($\mathrm{\mass}$), total transverse jet 
mass ($\mathrm{\massp}$) and total jet broadening ($\mathrm{\bT}$) and defined as ratios 
of momenta of the jets in an event and hence many uncertainties cancel out. These ESVs have 
higher values for multijet, spherical events and lower values for two-jet, pencil like events. 
The complement of transverse thrust ($\mathrm{\taup}$) is expected to be  sensitive to the 
initial hard scattering, multi-parton interaction, and emissions of high $\pT$ gluons 
in the form of ISR and FSR. The other three - $\mathrm{\mass}$,  $\mathrm{\massp}$ and 
$\mathrm{\bT}$ - have additional dependence on the details of fragmentation and hadronization. 
  
The distributions of the ESVs observed in data are unfolded to remove the effects of the 
efficiencies and acceptances of the CMS detector and a broad agreement between data and the 
predictions of \PY, \HWpp and \MGNLO is observed. In case of \PY, it is noted that the agreement 
is better for the $\mathrm{\taup}$ and $\mathrm{\massp}$ which are computed using $\pT$ of the jets. 
However, in case of $\mathrm{\bT}$ and $\mathrm{\mass}$, which are computed using $\vec{p}$ of 
the jets, \PY overestimates the multijet nature of the events. This indicates that the flow of 
energy in the transverse plane is better modelled by \PY while the overall three-dimensional 
modelling is not adequate.

CMS also observes \cite{Sirunyan:2018adt} that ISR has a large effect on the distributions of 
the ESVs and increases the spherical nature of the multijet events. The effect of FSR is 
much smaller and that of MPI is negligible.
  
For all the four ESVs, the overall agreement improves with the energy scale of the event. 
The strong coupling $\mathrm{\alpS}$ is smaller for higher energy and reduces 
the emission of hard gluons in the early stage of the evolution of the partons.  
This makes the higher order calculations less important for describing such 
event. CMS has used average $\pT$ of the two leading jets of an event, 
$\HT = (p_\textrm{T,\, jet1} + p_\textrm{T,\, jet2})/2$, to represent its energy scale.

\section{Optimizing the Parton Shower Model of PYTHIA8}\label{sec:4} 
Recently, both CMS and ATLAS have done several optimizations of \PY around its Monash 
tune \cite{Skands:2014pea}. CMS has used underlying event (UE) and charged particle data 
from the CDF ($\mathrm{p\bar{p}}$ collisions, Fermilab) and CMS (pp collisions, LHC) for 
tuning the strong coupling and MPI related parameters for different PDF sets 
\cite{Khachatryan:2015pea,Sirunyan:2019dfx}. ATLAS has optimized ISR, FSR and MPI related 
parameters using a number of observables \cite{Buckley:2014ctn}. The present study also 
uses the Monash tune, \PYTHIA v8.235 with NNPDF2.3 PDF (LO) set, as the default. For 
each parameter, variation has been done around the default value of the Monash tune. 

Since Monash tune overestimates the multijet nature of the events \cite{Sirunyan:2018adt}, 
the role of ISR and FSR need to be examined along with the choice of a suitable value of 
$\mathrm{\alpS}$. \PY has the provision to use separate values of $\mathrm{\alpS(M_{Z})}$ for ISR 
and FSR and the corresponding parameters are {\tt SpaceShower:alphaSvalue} and 
{\tt TimeShower:alphaSvalue} respectively. These will be referred to as $\mathrm{\asISR}$ 
and $\mathrm{\asFSR}$. The Monash tune uses $\mathrm{\alpS(M_Z) = 0.1365}$ 
for both. The maximum evolution scale involved in the showering is set to match the scale 
of the hard process itself. In \PY it set equal to the factorization scale,
but allows its modification by multiplicative factors {\tt SpaceShower:PTmaxFudge} for ISR 
and {\tt TimeShower:PTmaxFudge} for FSR. Both parameters in the Monash tune have 1 as their default values.

For each point in the parameter space, $10^{6}$ events have been generated to ensure that 
the statistical uncertainty is smaller than the experimental uncertainty. The resulting 
distributions have been compared with data and $\mathrm{\cindf}$ has been calculated to 
check the goodness of fit. 
The optimization has been done using \PROFESSOR v2.3.0 \cite{Buckley:2009bj} along with 
\RIVET v2.6 \cite{Buckley:2010ar} with the complete set of ESVs distributions from 
\PY as available from \cite{Sirunyan:2018adt}. Post optimization, the new parameter 
set is checked using other relevant results from the CMS \cite{Khachatryan:2016wdh} and 
ATLAS \cite{ATLAS:2016djc}. 

\subsection{Optimization of the Strong Coupling for ISR and FSR}
The Monash tune of \PY uses the same value of $\mathrm{\alpS}$ for both ISR and FSR as 
the default. It is a choice of convenience but not mandatory. It has been argued that the 
effective scales for ISR and FSR are different which could be absorbed by effective values of 
the corresponding strong coupling and this is acceptable. This freedom to set nonidentical 
$\mathrm{\alpS}$ for ISR and FSR within the validity of the PS approach gives MC tuning studies 
a space to best describe the observables. The observation of CMS \cite{Sirunyan:2018adt} that 
ISR has a rather large effect on the ESVs compared to FSR adds to the importance of this issue. 
So it is instructive and important to probe if \PY prefers to have different values of 
$\mathrm{\alpS}$ for both ISR and FSR.

\begin{figure}[tbp]
	\centering
   \includegraphics[page=1, width=5.5cm]{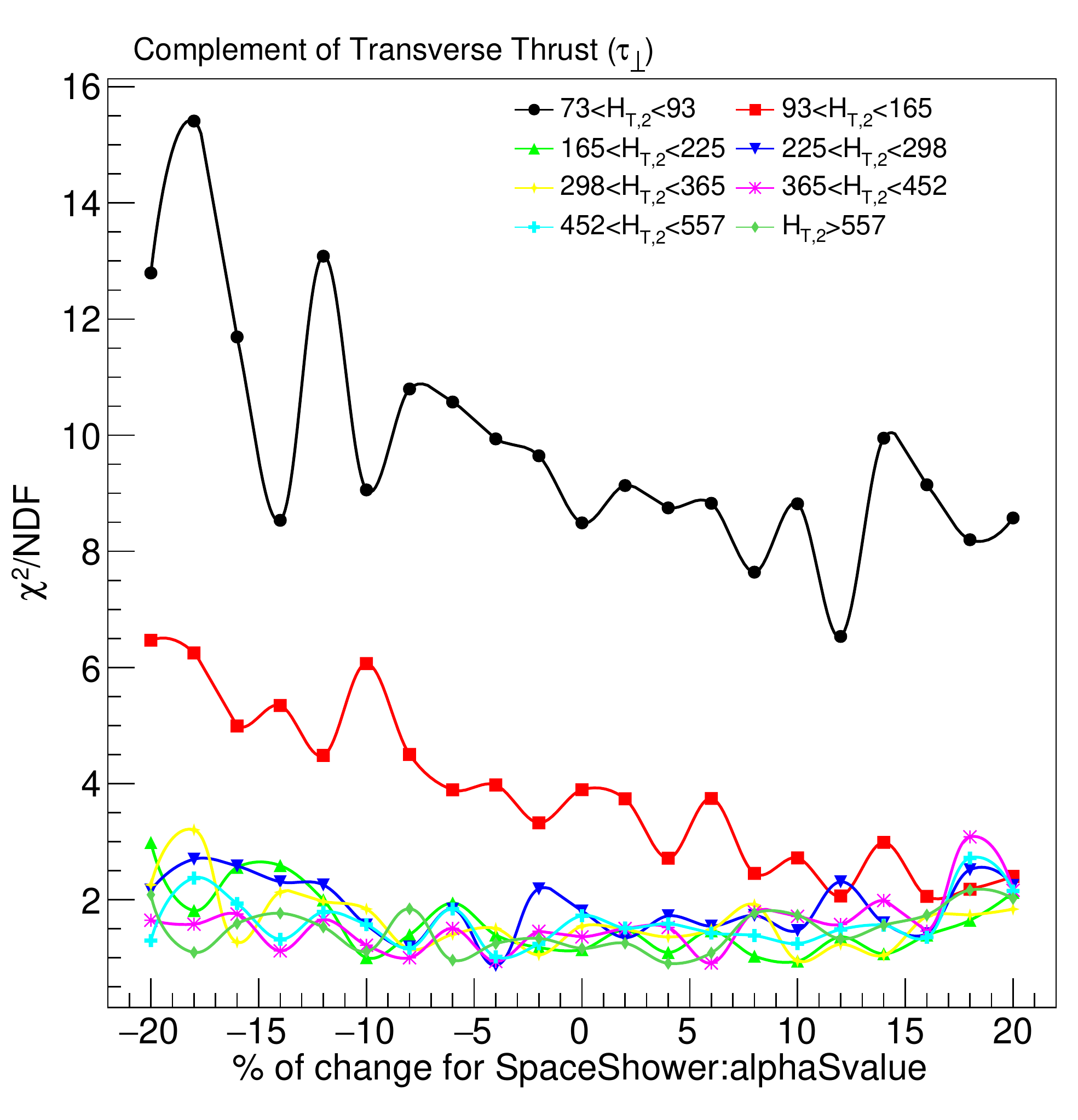}
   \includegraphics[page=2, width=5.5cm]{CHI2NDF_ISR_20_20_Final.pdf}
   \includegraphics[page=3, width=5.5cm]{CHI2NDF_ISR_20_20_Final.pdf}
   \includegraphics[page=4, width=5.5cm]{CHI2NDF_ISR_20_20_Final.pdf}
   \caption{Variation of $\mathrm{\cindf}$ with {\tt SpaceShower:alphaSvalue} ($\asISR$) used in ISR,
            for the ESVs - the complement of transverse thrust (top left), total
            transverse jet mass(top right), total jet mass (bottom left), and
            total jet broadening (bottom right). The axes represent variation with
            respect to the default values used in the Monash tune of \PY.}
\label{Fig:ISR_CHI1}
\end{figure}

First, ~$\mathrm{\asISR}$ is varied, keeping ~$\mathrm{\asFSR}$ and other parameters fixed at 
their default values. The value of $\asISR$ is varied by 20\% about the default value in steps 
of 2\%. The results, in terms of $\mathrm{\cindf}$ are shown in figure \ref{Fig:ISR_CHI1}. 
It is seen that $\mathrm{\taup}$ and $\mathrm{\massp}$ are not sensitive to $\asISR$ while 
$\mathrm{\mass}$ and $\mathrm{\bT}$ is mildly sensitive preferring a lower value. 

\begin{figure}[tbp]
	\centering
   \includegraphics[page=1, width=5.5cm]{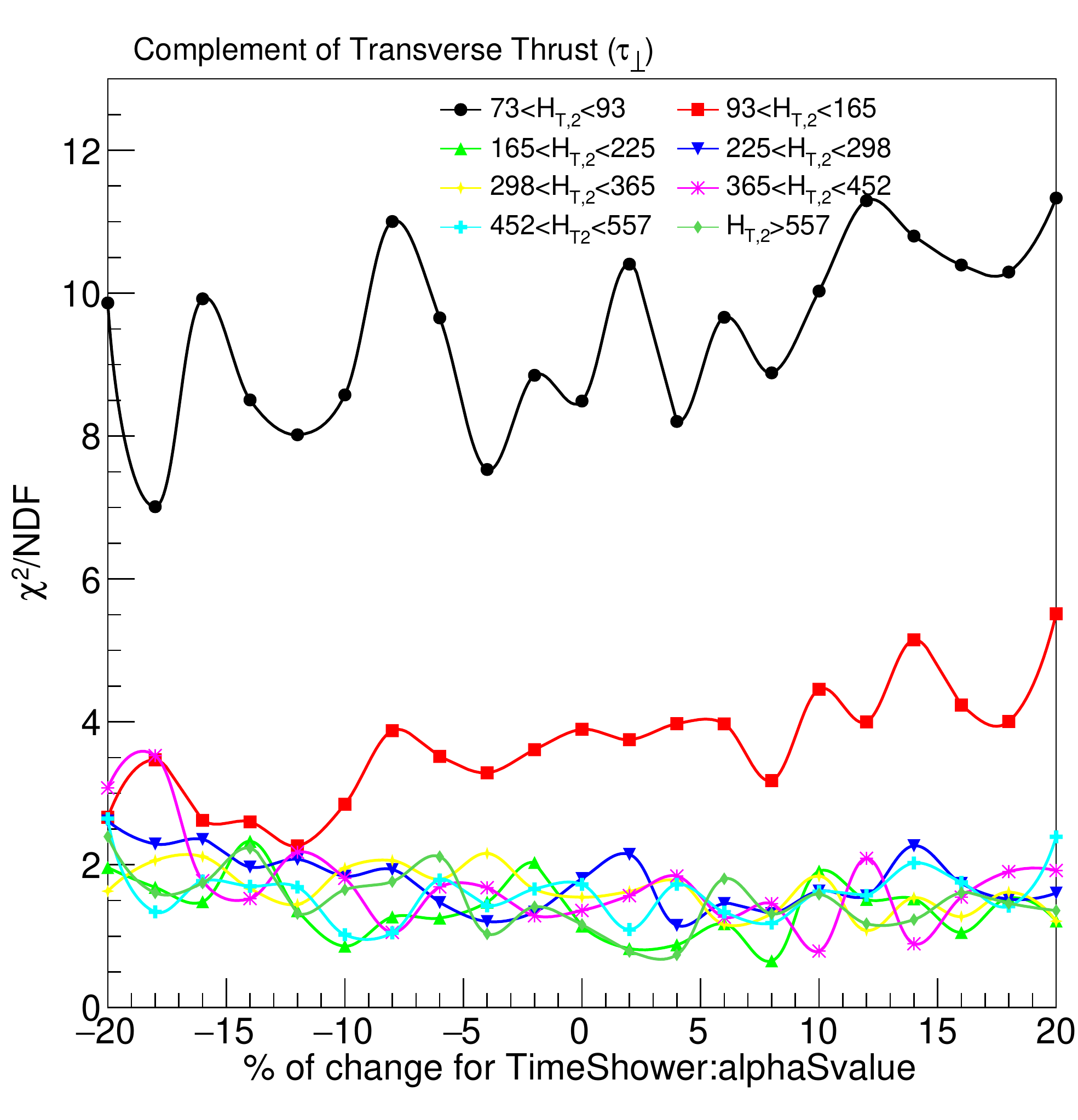}
   \includegraphics[page=2, width=5.5cm]{CHI2NDF_FSR_20_20_Final.pdf}
   \includegraphics[page=3, width=5.5cm]{CHI2NDF_FSR_20_20_Final.pdf}
   \includegraphics[page=4, width=5.5cm]{CHI2NDF_FSR_20_20_Final.pdf}
   \caption{Variation of $\mathrm{\cindf}$ with {\tt TimeShower:alphaSvalue} ($\asFSR$) used in FSR,
            for the ESVs - the complement of transverse thrust (top left), total 
            transverse jet mass(top right), total jet mass (bottom left), and 
            total jet broadening (bottom right). The axes represent variation with 
            respect to the default values used in the Monash tune of \PY.}
 \label{Fig:FSR_CHI1}
\end{figure}

Next, $\mathrm{\asFSR}$ is varied, keeping other parameters including $\mathrm{\asISR}$ fixed at 
their default values. The value of $\mathrm{\asFSR}$ is also varied by 20\% in steps of 
2\%. The results, in terms of $\mathrm{\cindf}$ are shown in figure \ref{Fig:FSR_CHI1}. 
It is seen that $\mathrm{\taup}$ is not sensitive to $\mathrm{\asFSR}$, $\mathrm{\massp}$ prefers 
the default value, $\mathrm{\mass}$, and $\mathrm{\bT}$ prefer a higher value. 

\begin{figure}[tbp]
	\centering
   \includegraphics[page=1, width=5.5cm]{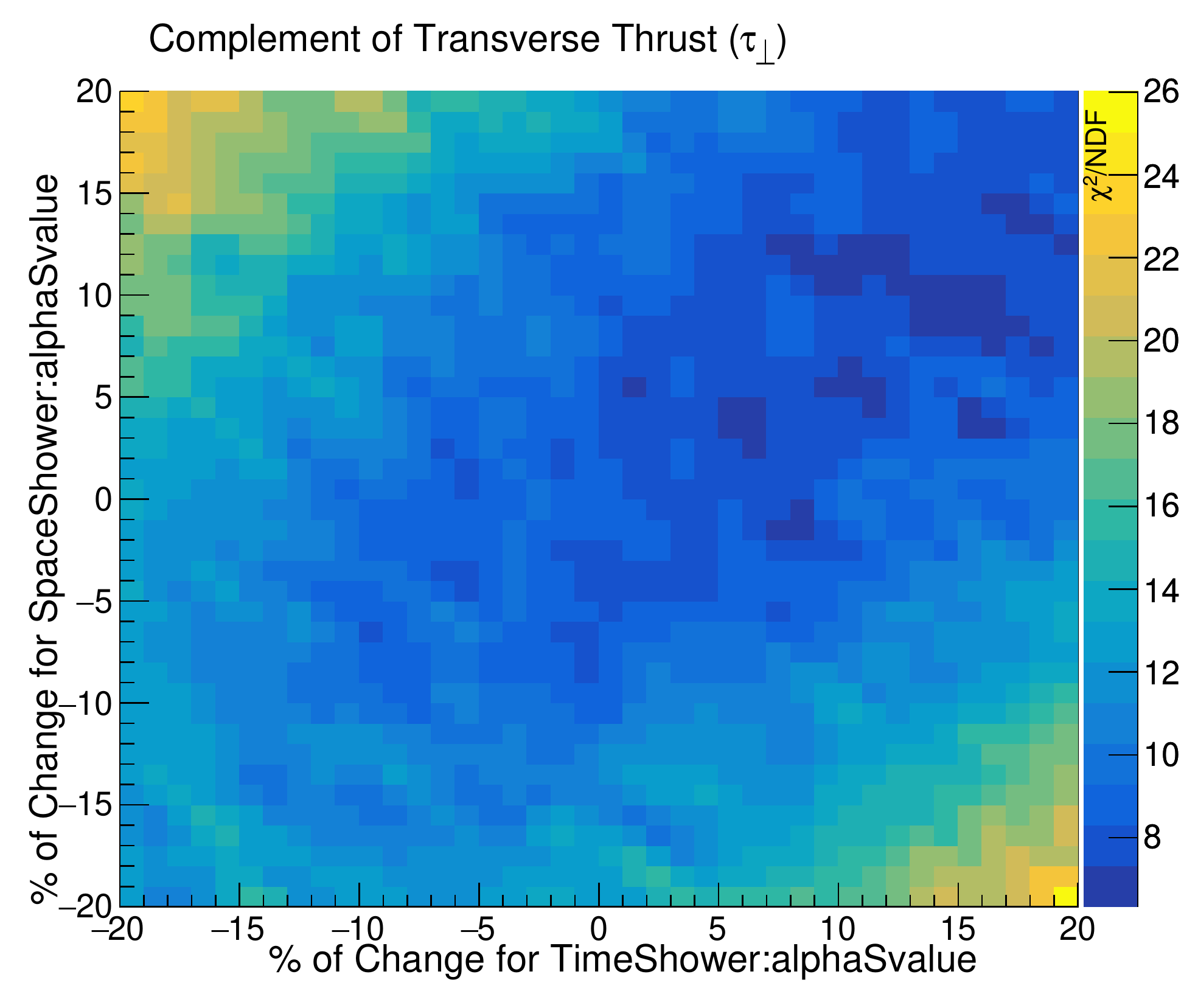}
   \includegraphics[page=2, width=5.5cm]{CHI2NDF_HT2Merge_Final.pdf}
   \includegraphics[page=3, width=5.5cm]{CHI2NDF_HT2Merge_Final.pdf}
   \includegraphics[page=4, width=5.5cm]{CHI2NDF_HT2Merge_Final.pdf}
        \caption{ Variation of $\mathrm{\chi^2/NDF}$, summed over six $\HT$ 
        ranges, with $\asISR$ and $\asFSR$ for the four ESVs, the complement of transverse thrust (top left), total 
        transverse jet mass(top right), total jet mass (bottom left), and total 
        jet broadening (bottom right). The axes represent variation with 
        respect to the default values used in the Monash tune of \PY.}
\label{Fig:ISR_FSR_COLL1}
\end{figure}

The effect of variation of either $\asISR$, or, $\asFSR$ on the four ESVs conform to the 
observations of CMS regarding the effects of ISR and FSR. The disagreements for the first two 
$\HT$ ranges for all of them are very large and appear to be beyond the scope of 
parameter tuning and left out of the next study, where both $\asISR$ and $\asFSR$ are varied 
simultaneously and over the same ranges as above. Each plot in figure \ref{Fig:ISR_FSR_COLL1} 
shows the summed up $\mathrm{\cindf}$ for the six $\HT$ ranges for the same variable. 
This gives us an overall idea of the preferable values of $\asISR$ and $\asFSR$.  

\subsection{Optimization of Maximum scale of the Shower Evolution}
\begin{figure}[tbp]
	\centering
   \includegraphics[page=1, width=5.5cm]{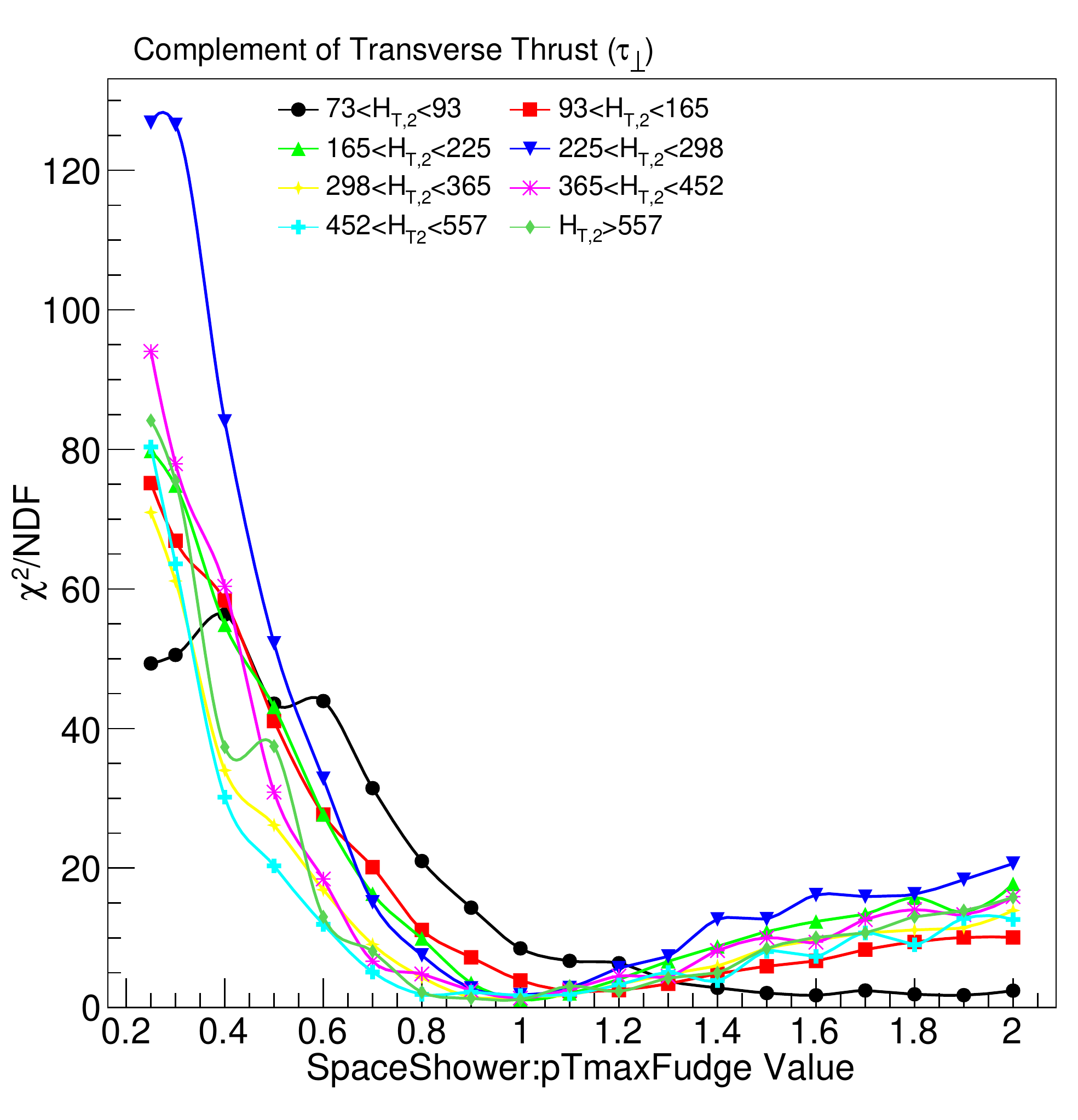}
   \includegraphics[page=2, width=5.5cm]{CHI2NDF_PTmaxFudge_Final.pdf}
   \includegraphics[page=3, width=5.5cm]{CHI2NDF_PTmaxFudge_Final.pdf}
   \includegraphics[page=4, width=5.5cm]{CHI2NDF_PTmaxFudge_Final.pdf}
        \caption{ Variation of $\mathrm{\chi^2/NDF}$ with {\tt SpaceShower:PTmaxFudge} 
        for the four ESVs, the complement of transverse thrust (top, left), total 
        transverse jet mass(top, right), total jet mass (bottom left), and total 
        jet broadening (bottom right).}
\label{Fig:Fudge_CHI2}
\end{figure}

The sensitivity of the ESVs to maximum scale of the ISR shower evolution {\tt SpaceShower:PTmaxFudge} 
($\mathrm{\tt PTmaxFudge^{ISR}}$) has been checked. Figure \ref{Fig:Fudge_CHI2} 
shows the variation of $\mathrm{\cindf}$ for the four ESVs with this parameter. 
It is observed that $\mathrm{\taup}$ and $\mathrm{\massp}$ have similar dependence on 
$\mathrm{\tt PTmaxFudge^{ISR}}$, preferring a value of $\mathrm{\sim 1}$. On the other hand, 
$\mathrm{\mass}$ and $\mathrm{\bT}$ prefer a value $\mathrm{\sim 0.6}$. Similar study 
for {\tt TimeShower:PTmaxFudge}, related to FSR, shows no significant variation for the ESVs.

\begin{table}[tbp]
\centering
\setlength\tabcolsep{11pt}
\begin{tabular}{|cccc|}
\hline   \PY &   Monash & Sampling range &  Optimized  \\
                Parameters set & values &   & values   \\
        \hline   {\tt SpaceShower:alphaSvalue}& 0.1365 & $0.1092-0.1638$ & $0.11409^{+0.00078}_{-0.00073}$  \\
           {\tt TimeShower:alphaSvalue}& 0.1365 & $0.1092-0.1638 $ &  $0.15052^{+0.00077}_{-0.00076}$  \\
           {\tt SpaceShower:PTmaxFudge}& 1.0 & $0.6-1.4$ &  $0.9323^{+0.0065}_{-0.0064}$  \\
           \hline
\end{tabular}
\caption{Optimization of three parameters of \PY is shown along with their default values in the Monash
tune and the sampling range.}
\label{tab:Param1}
\end{table}


\subsection{Simultaneous Optimization of the Strong Coupling and Maximum scale of the Shower Evolution}
Finally, \PROFESSOR v2.3.0 \cite{Buckley:2009bj} is used to simultaneously optimize 
$\mathrm{\asISR}$, $\mathrm{\asFSR}$, and $\mathrm{\tt PTmaxFudge^{ISR}}$ keeping other parameters fixed to their 
default values in the Monash tune. CMS data for the four ESVs are made available through 
{\RIVET} v2.6 \cite{Buckley:2010ar} and HEP Data~\cite{1701612}. Overall, 120 different 
combinations of the three parameters are randomly sampled by \PROFESSOR in the ranges  
mentioned in table \ref{tab:Param1}. \PROFESSOR was instructed to perform a third order 
polynomial fit to optimize  parameter response 
for the observables, the four ESVs in this case. This goodness-of-fit minimization process  
gives in return best favourable values of the three parameters as listed in table \ref{tab:Param1}. 
The uncertainty for the optimized parameters are obtained from Minuit~\cite{james1975minuit} package 
interfaced with \PROFESSOR. 

The values obtained for $\mathrm{\asISR}$, $\mathrm{\asFSR}$, and $\mathrm{\tt PTmaxFudge^{ISR}}$ are 
0.11409, 0.15052 and 0.9323 respectively, see table \ref{tab:Param1}. It is noted that the model of QCD 
implemented in \PYTHIA prefers a lower value of $\mathrm{\alpS}$ for ISR compared to the default 
value used in the Monash tune but prefers a higher value of the same for FSR. It is also observed that 
the optimized values obtained here for $\mathrm{\asISR}$ and $\mathrm{\asFSR}$ are very close to their values
obtained while $\mathrm{\tt PTmaxFudge^{ISR}}$ is fixed to 1.
\begin{figure}[tbp]
\centering
  \includegraphics[page=1, width=.38\textwidth, height =.37\textwidth ]{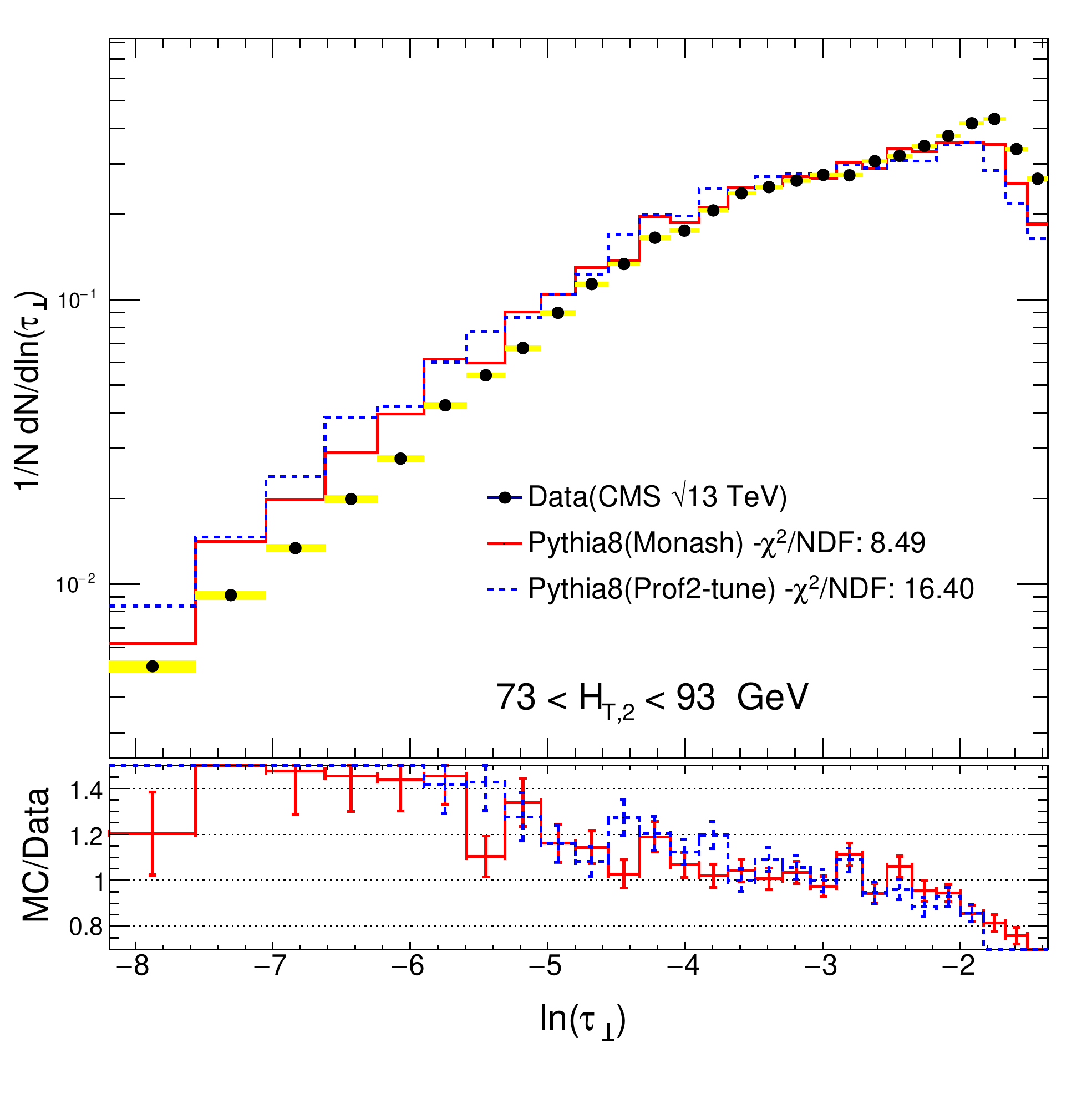}
  \hspace{.5cm}
  \includegraphics[page=2, width=.38\textwidth, height =.37\textwidth ]{ESV_Monash_Prof2_Root_v1.pdf}
  \includegraphics[page=3, width=.38\textwidth, height =.37\textwidth ]{ESV_Monash_Prof2_Root_v1.pdf}
    \hspace{.5cm}
  \includegraphics[page=4, width=.38\textwidth, height =.37\textwidth ]{ESV_Monash_Prof2_Root_v1.pdf}
  \includegraphics[page=5, width=.38\textwidth, height =.37\textwidth ]{ESV_Monash_Prof2_Root_v1.pdf}
  \hspace{.5cm}
  \includegraphics[page=6, width=.38\textwidth, height =.37\textwidth ]{ESV_Monash_Prof2_Root_v1.pdf}
  \includegraphics[page=7, width=.38\textwidth, height =.37\textwidth ]{ESV_Monash_Prof2_Root_v1.pdf}
  \hspace{.5cm}
  \includegraphics[page=8, width=.38\textwidth, height =.37\textwidth ]{ESV_Monash_Prof2_Root_v1.pdf}
\caption{\PY with the optimized parameter set is compared with CMS data. Monash tune of \PY 
is also shown for comparison. The plots show normalized distributions of the complement of transverse 
thrust($\mathrm{\taup}$) for eight $\HT$ ranges and the bottom panel in each plot shows 
ratios of MC with data.}
\label{Fig:Monash_taup}
\end{figure}

\begin{figure}[tbp]
\centering
  \includegraphics[page=9, width=.38\textwidth, height =.37\textwidth ]{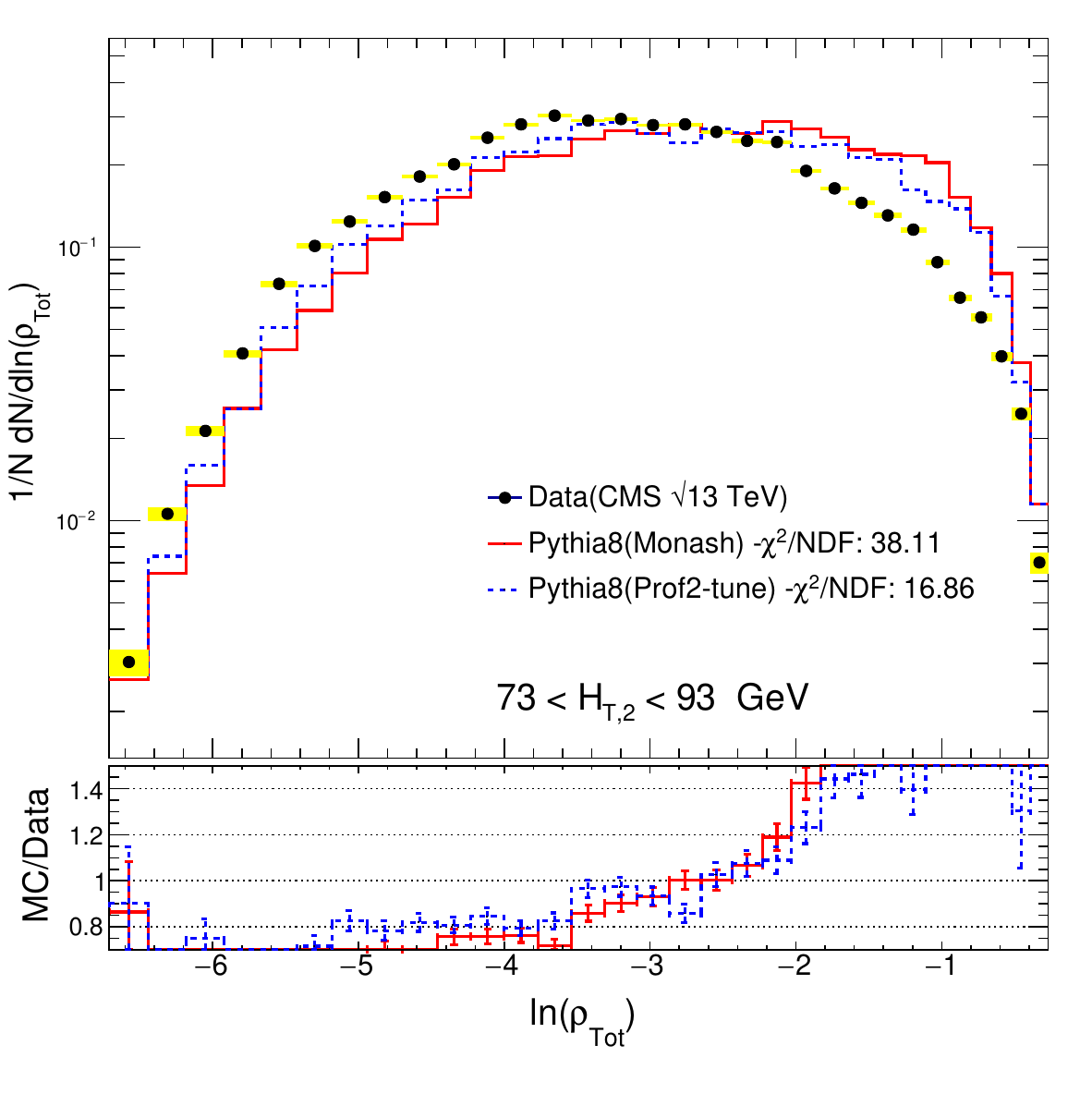}
  \hspace{.5cm}
  \includegraphics[page=10, width=.38\textwidth, height =.37\textwidth ]{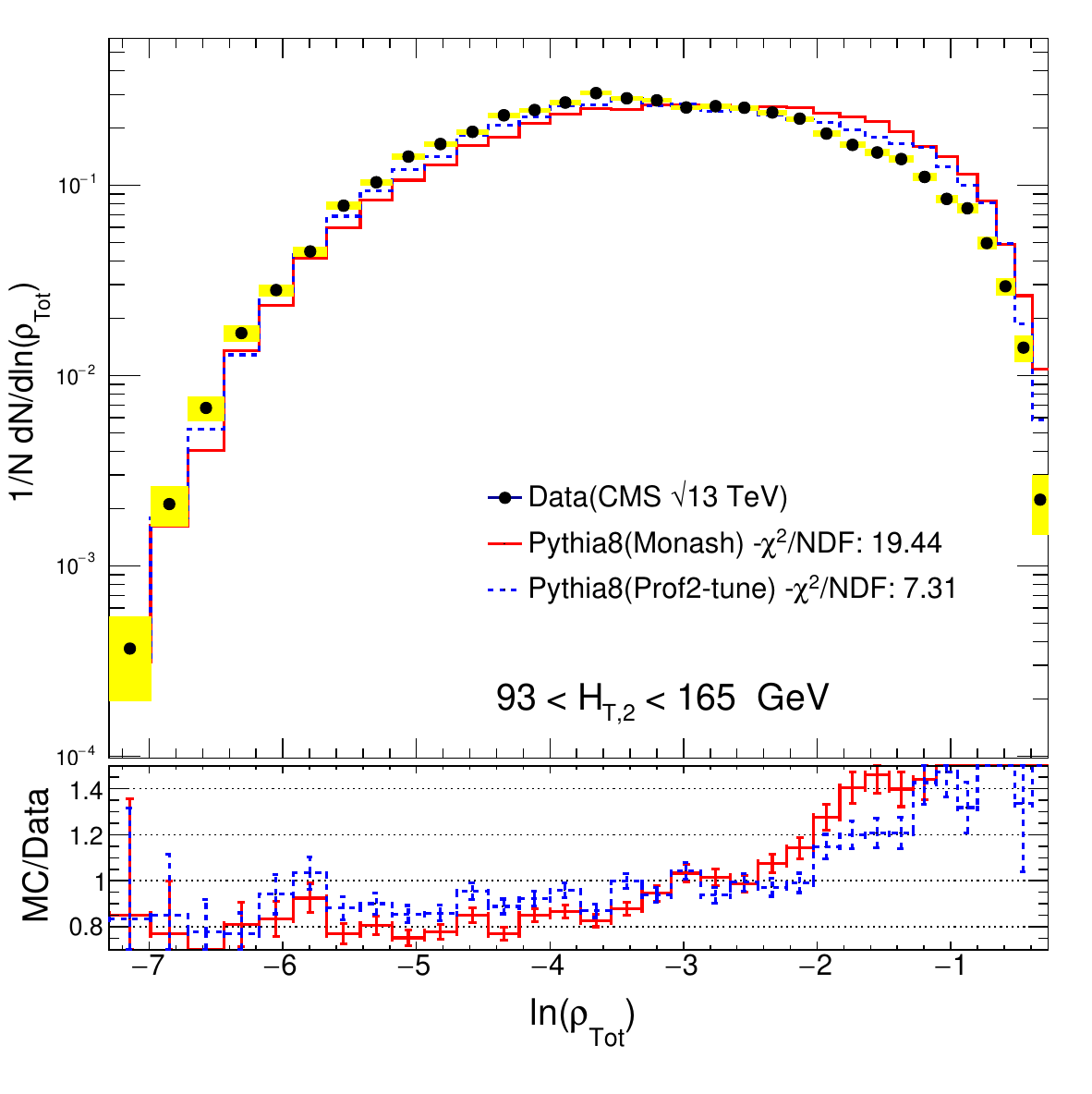}
  \includegraphics[page=11, width=.38\textwidth, height =.37\textwidth ]{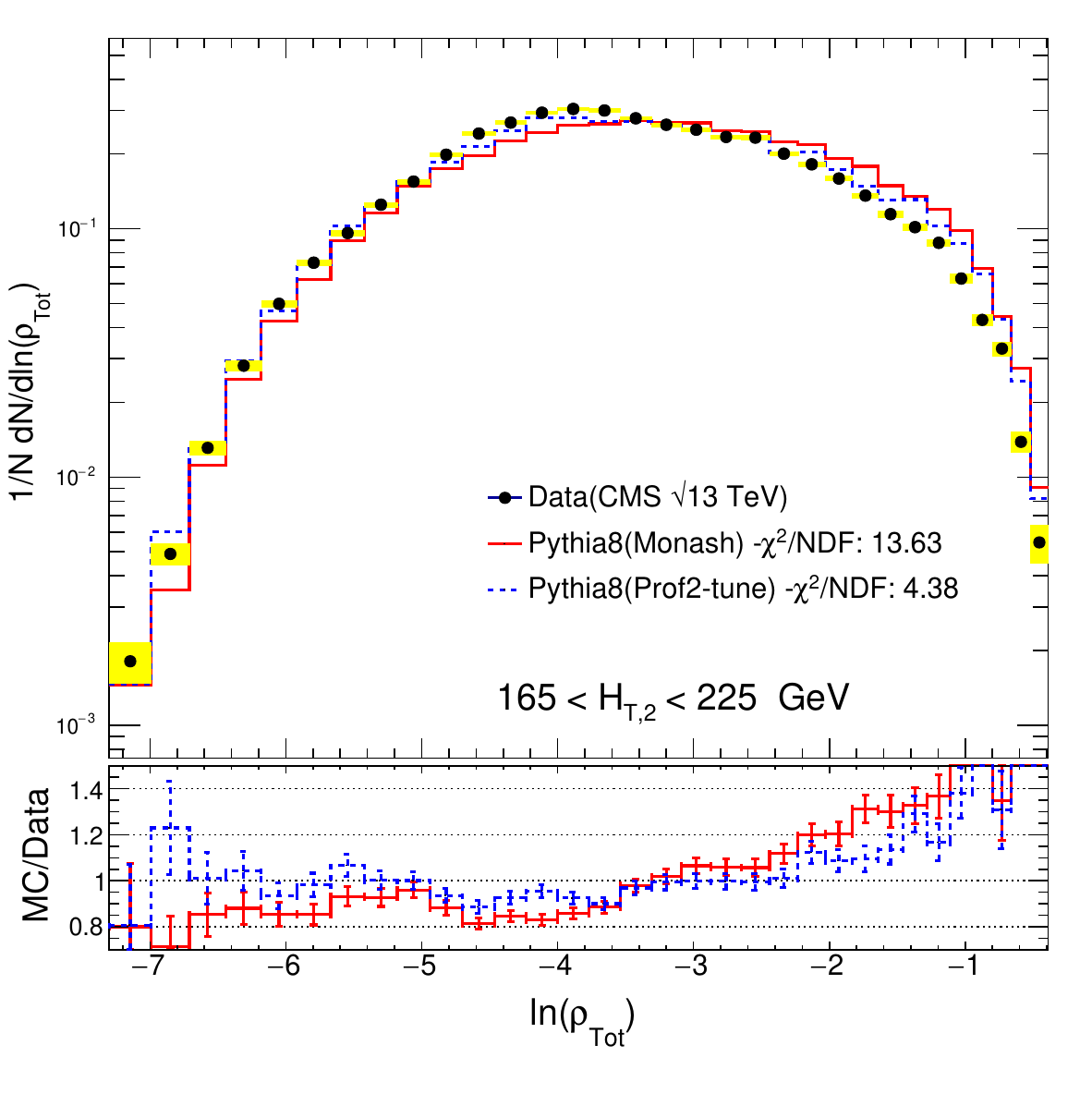}
  \hspace{.5cm}
  \includegraphics[page=12, width=.38\textwidth, height =.37\textwidth ]{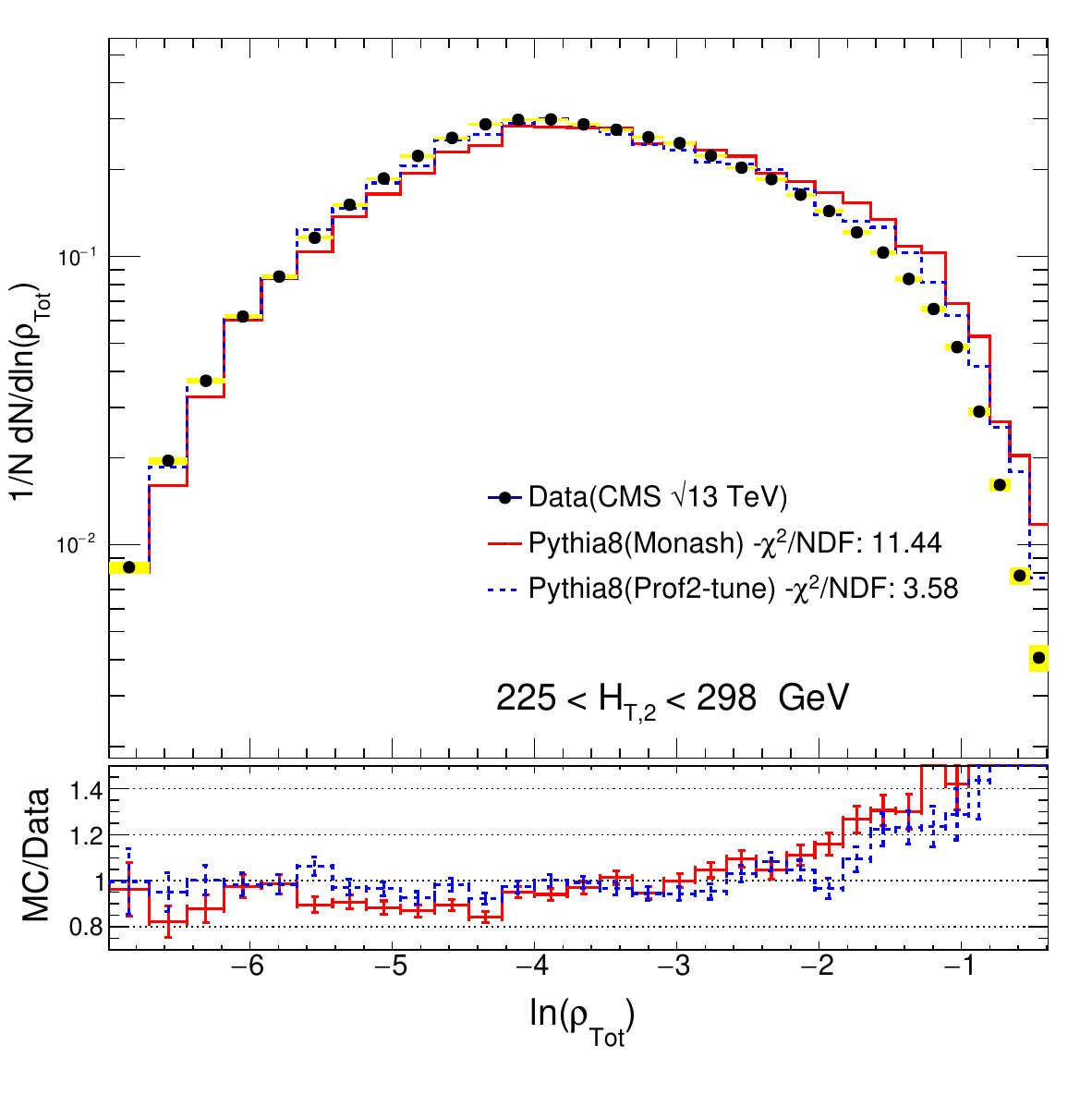}
  \includegraphics[page=13, width=.38\textwidth, height =.37\textwidth ]{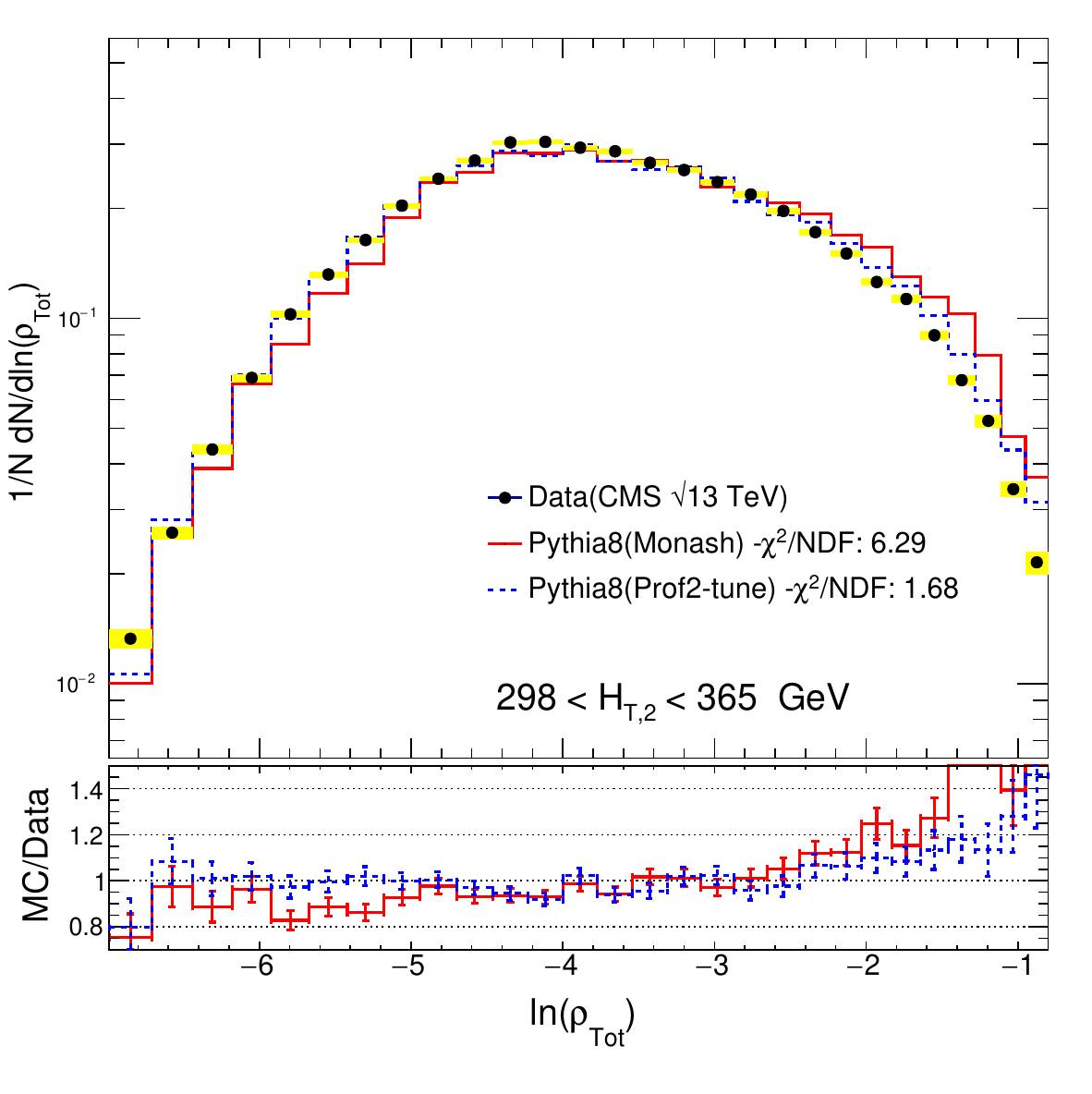}
  \hspace{.5cm}
  \includegraphics[page=14, width=.38\textwidth, height =.37\textwidth ]{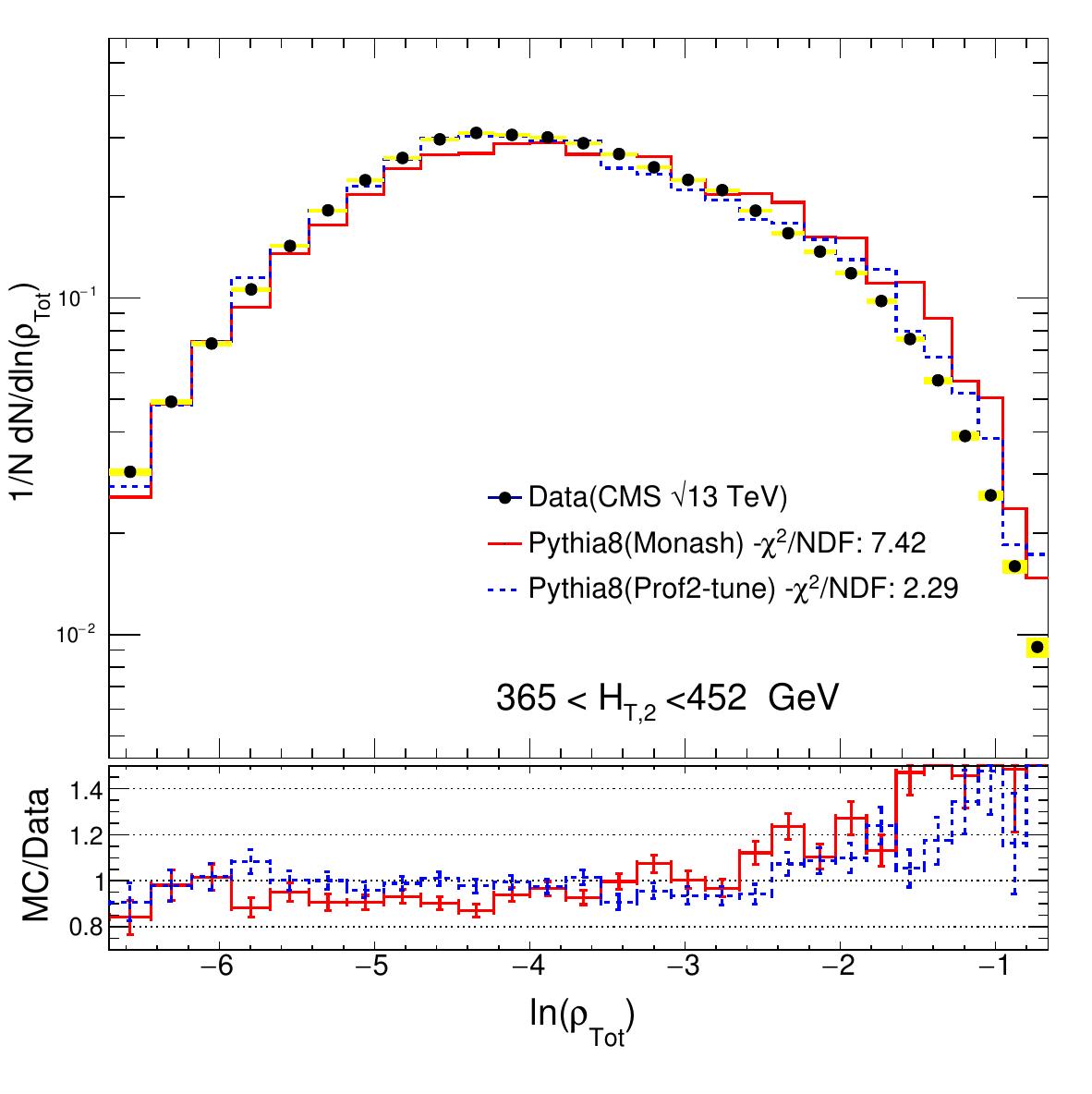}
  \includegraphics[page=15, width=.38\textwidth, height =.37\textwidth ]{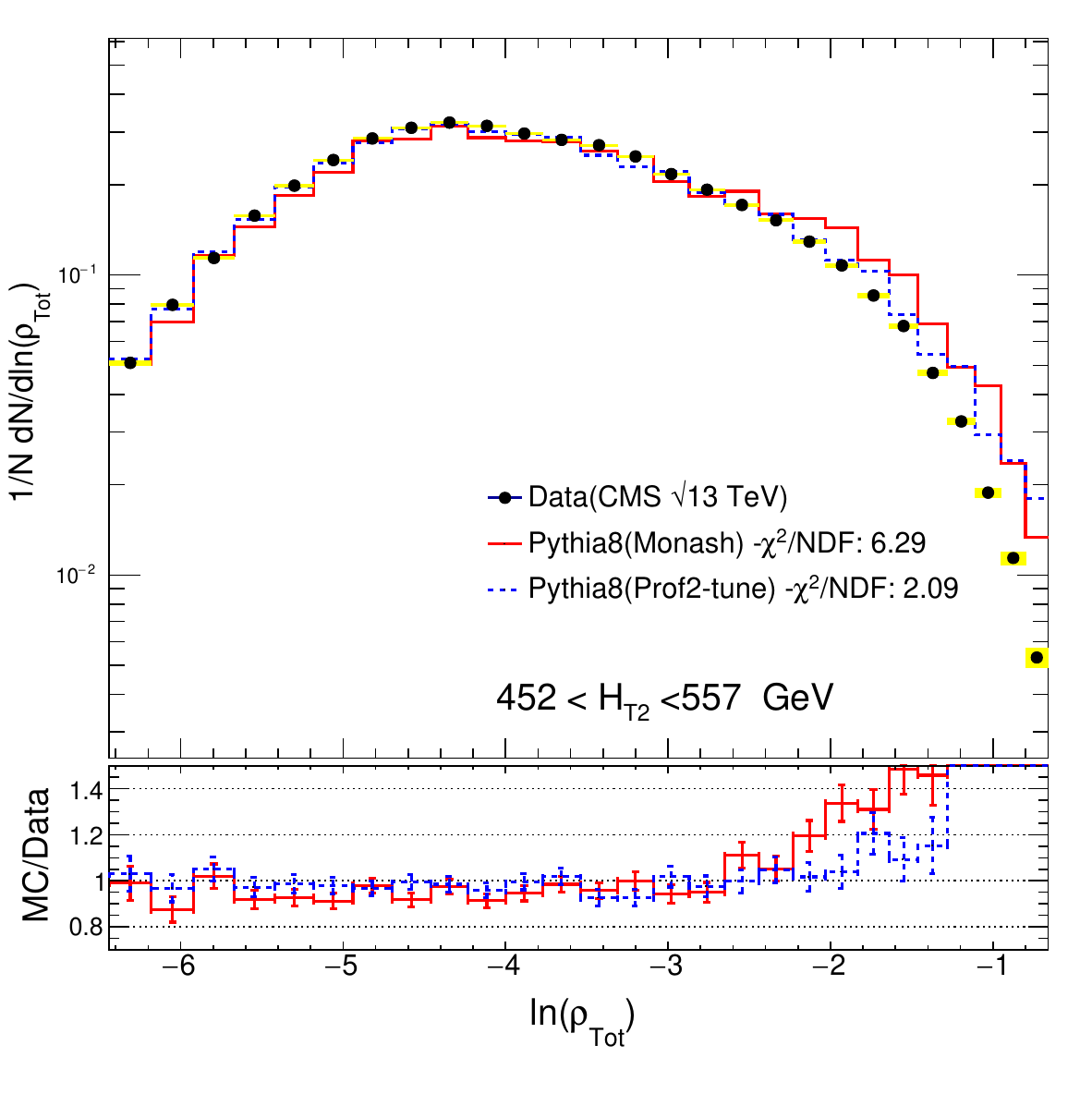}
  \hspace{.5cm}
  \includegraphics[page=16, width=.38\textwidth, height =.37\textwidth ]{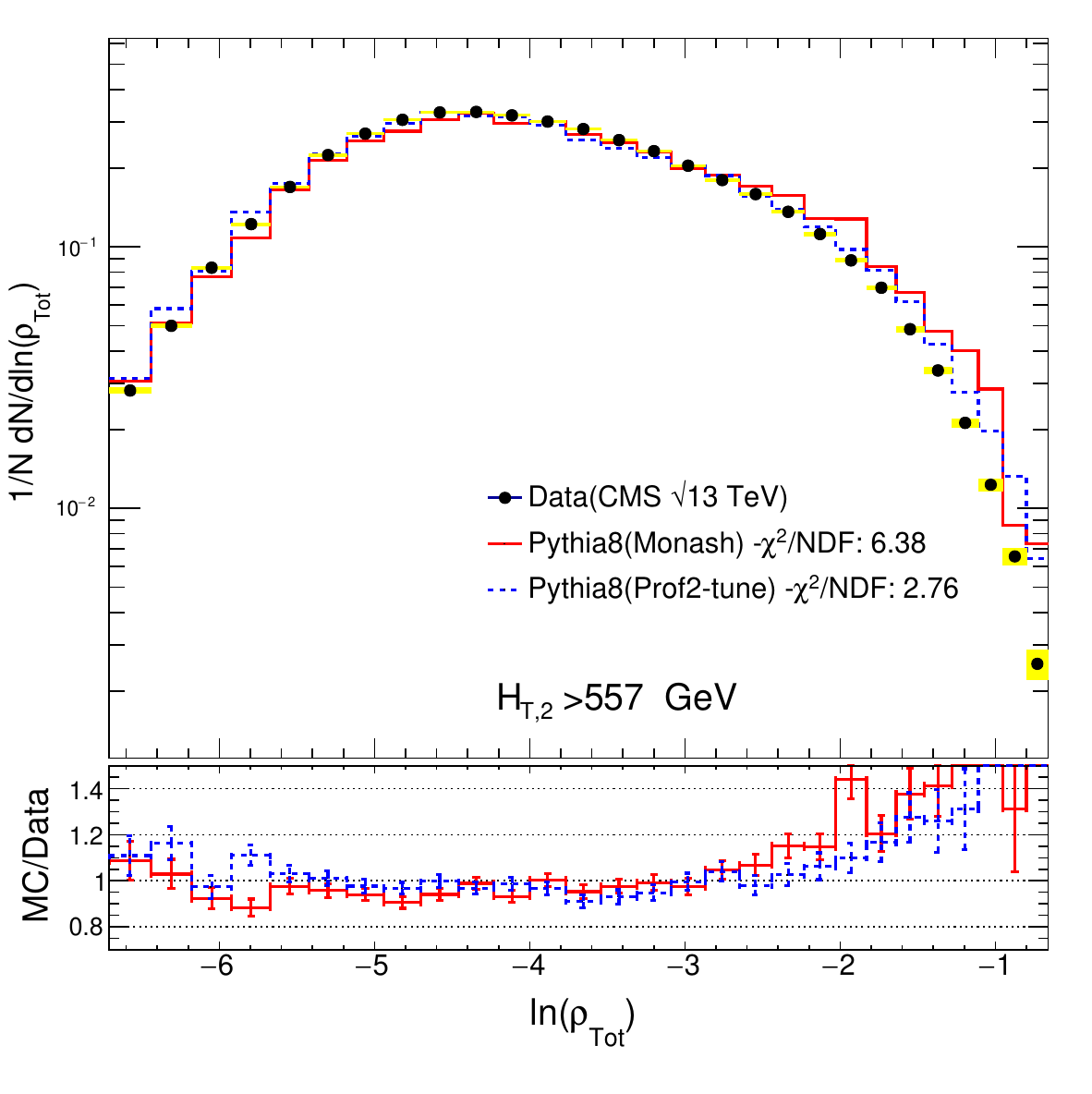}
\caption{\PY with the optimized parameter set is compared with CMS data. Monash tune of \PY 
is also shown for comparison. The plots show normalized distributions of the total jet mass 
($\mathrm{\mass}$) for eight $\HT$ ranges and the bottom panel in each plot shows 
ratios of MC with data.}
\label{Fig:Monash_mass}
\end{figure}

\begin{figure}[tbp]
\centering
  \includegraphics[page=17, width=.38\textwidth, height =.37\textwidth ]{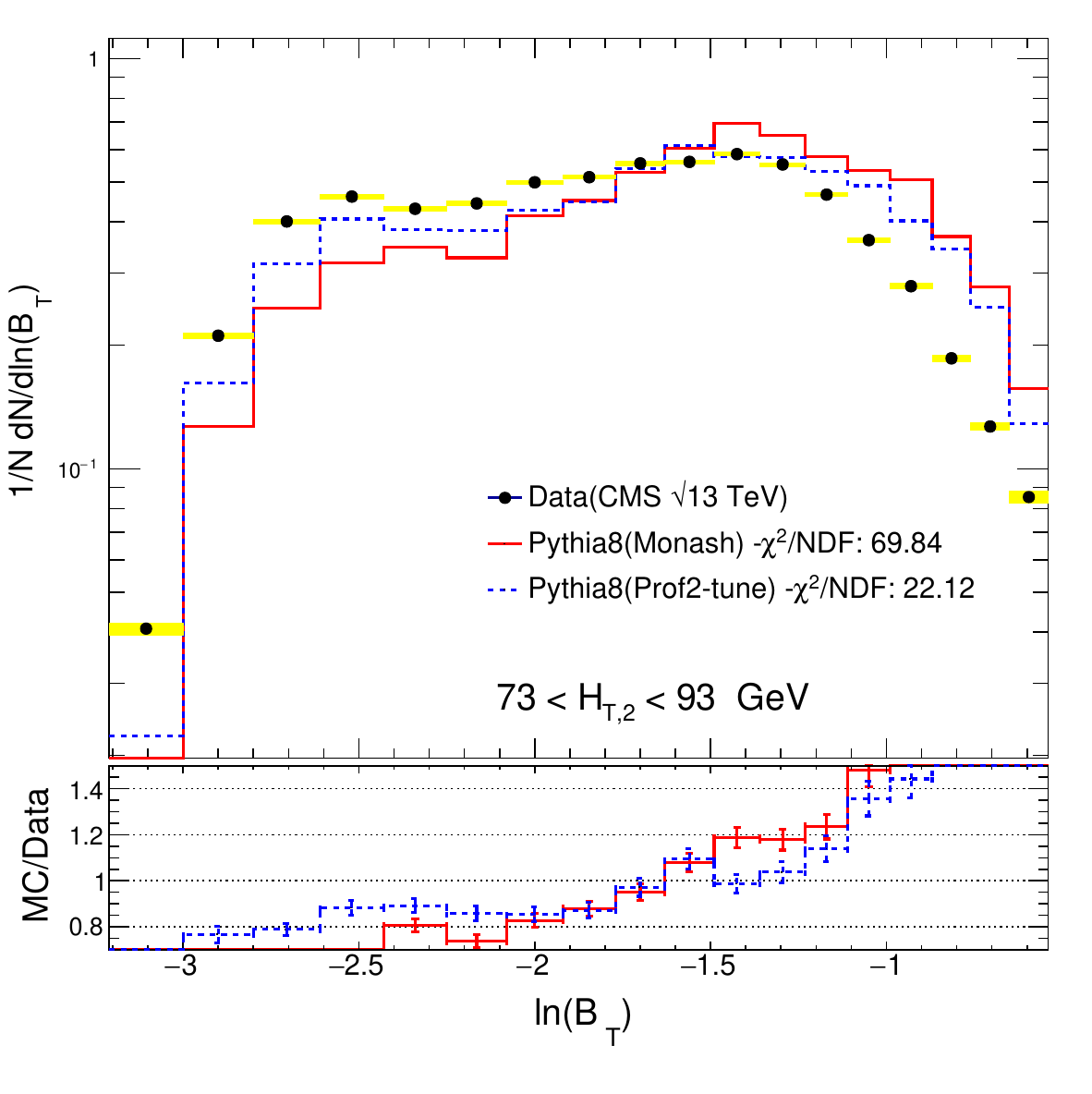}
  \hspace{.5cm}
  \includegraphics[page=18, width=.38\textwidth, height =.37\textwidth ]{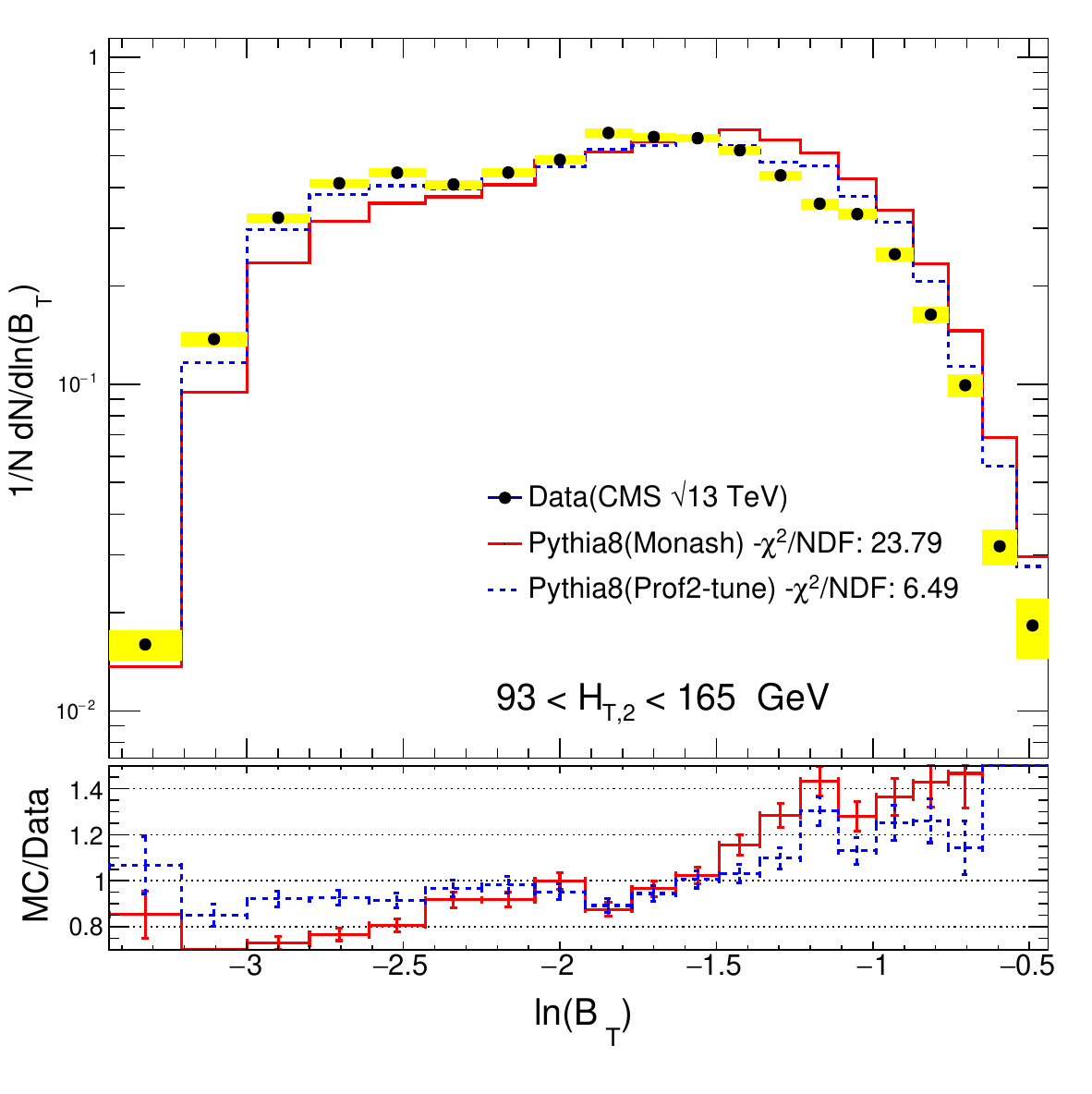}
  \includegraphics[page=19, width=.38\textwidth, height =.37\textwidth ]{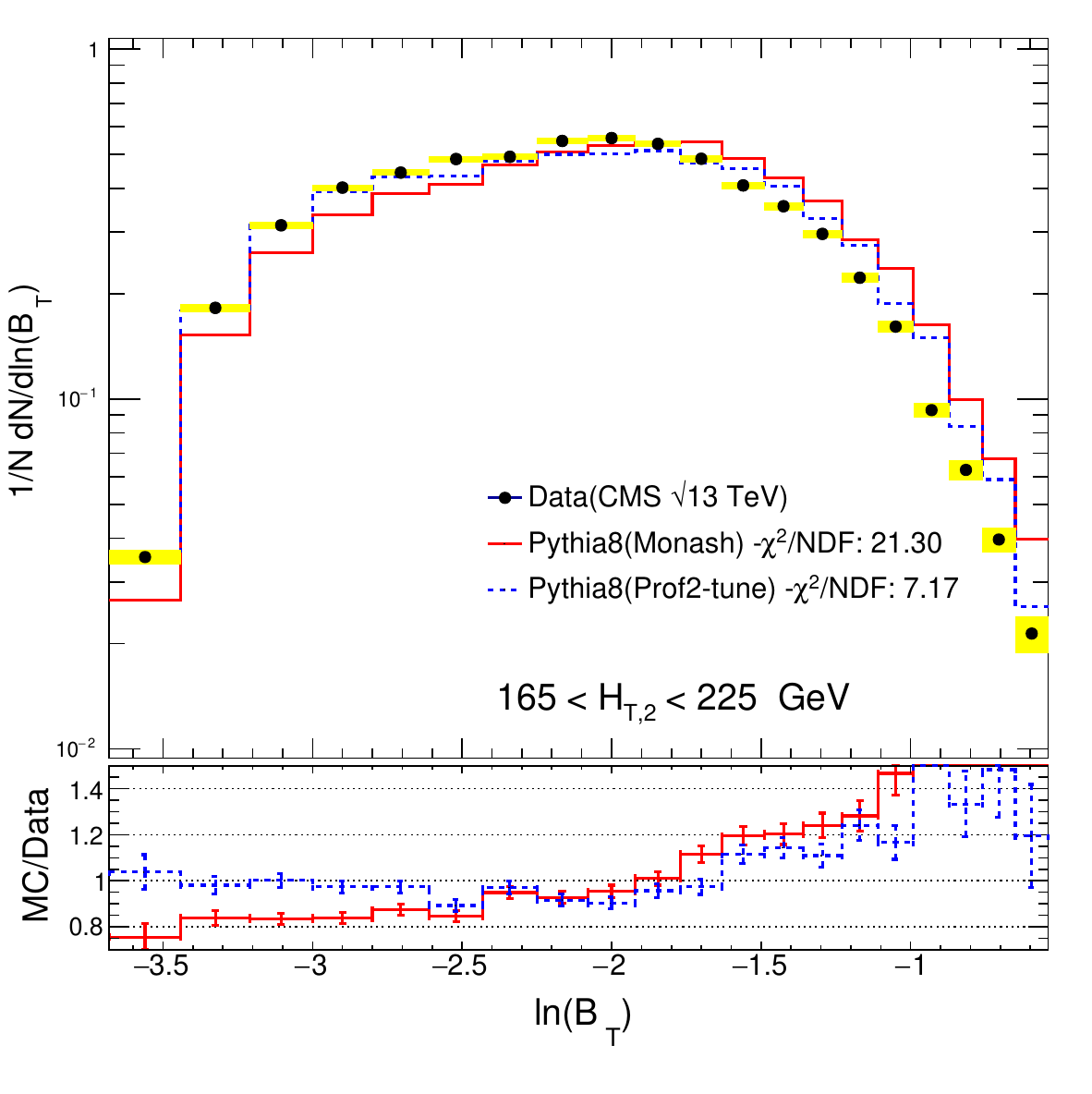}
  \hspace{.5cm}
  \includegraphics[page=20, width=.38\textwidth, height =.37\textwidth ]{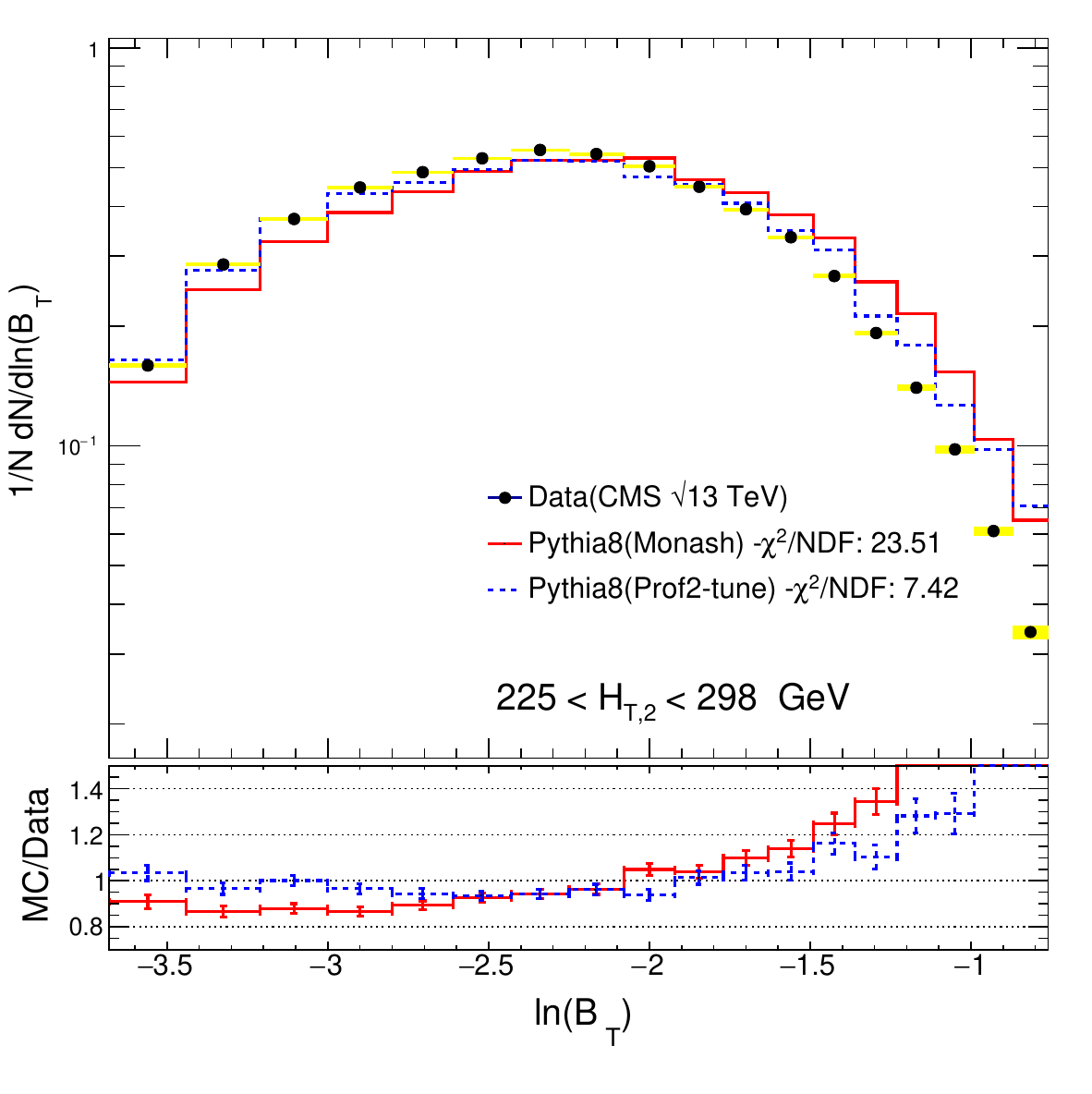}
  \includegraphics[page=21, width=.38\textwidth, height =.37\textwidth ]{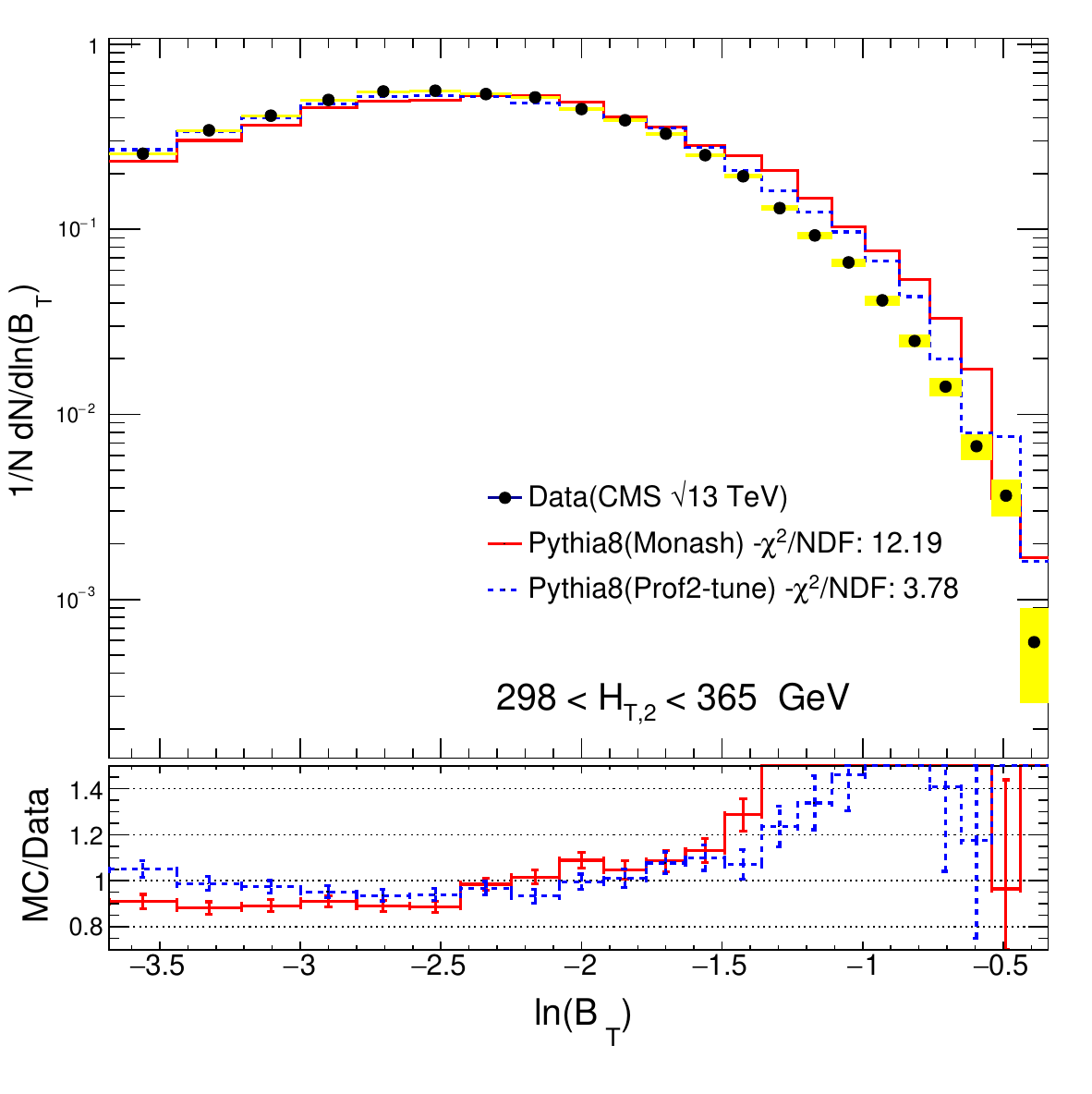}
  \hspace{.5cm}
  \includegraphics[page=22, width=.38\textwidth, height =.37\textwidth ]{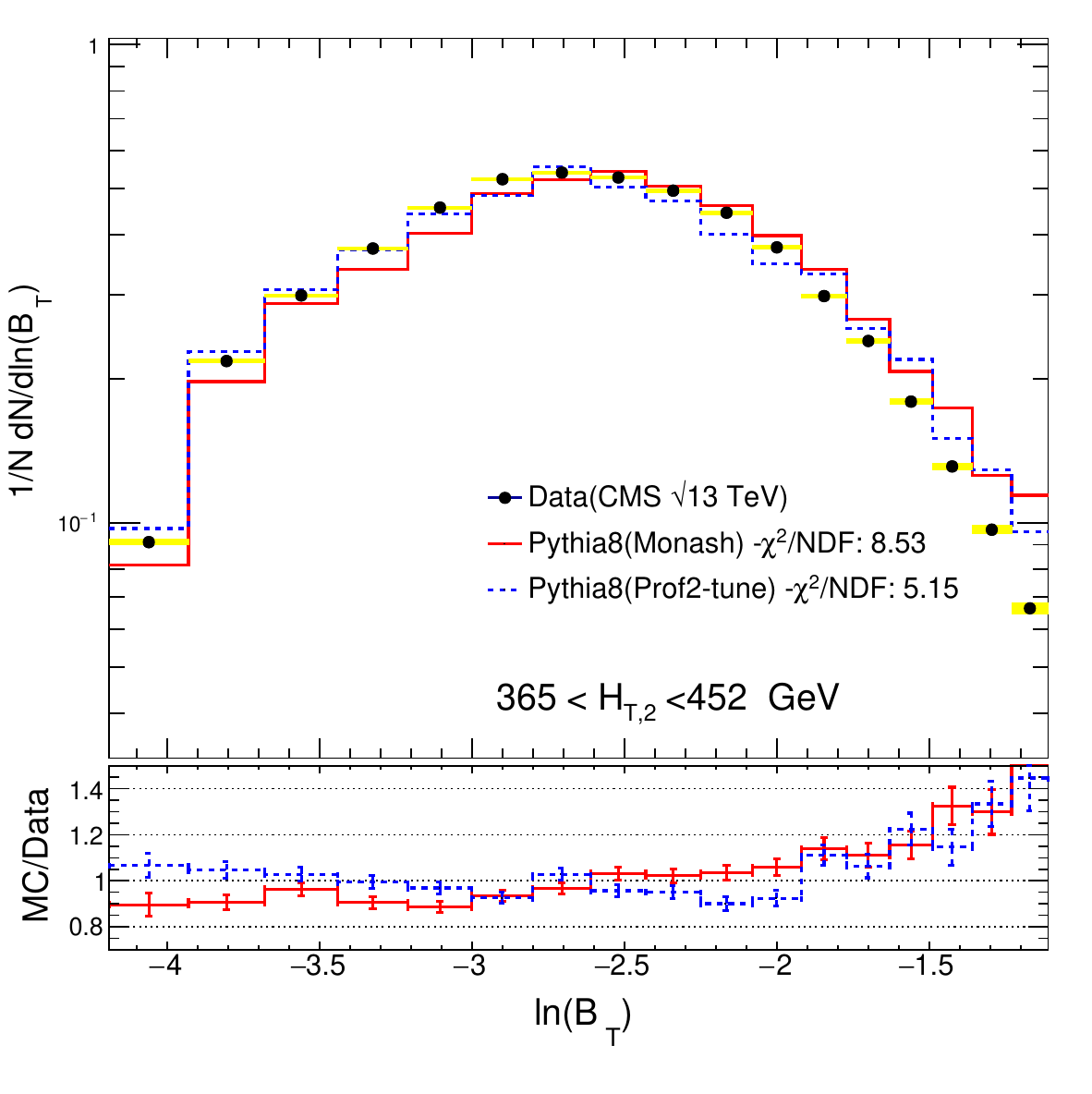}
  \includegraphics[page=23, width=.38\textwidth, height =.37\textwidth ]{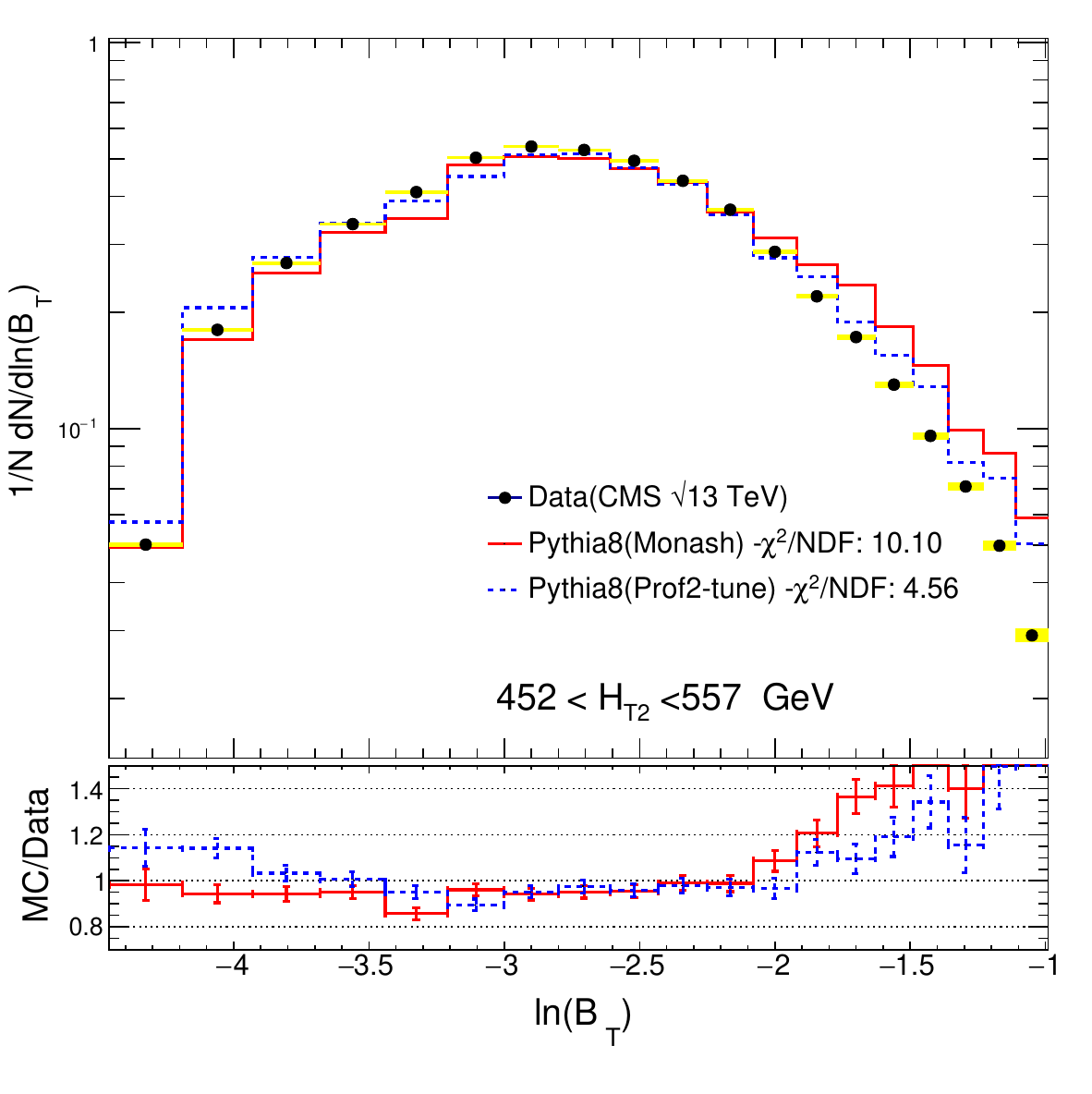}
  \hspace{.5cm}
  \includegraphics[page=24, width=.38\textwidth, height =.37\textwidth ]{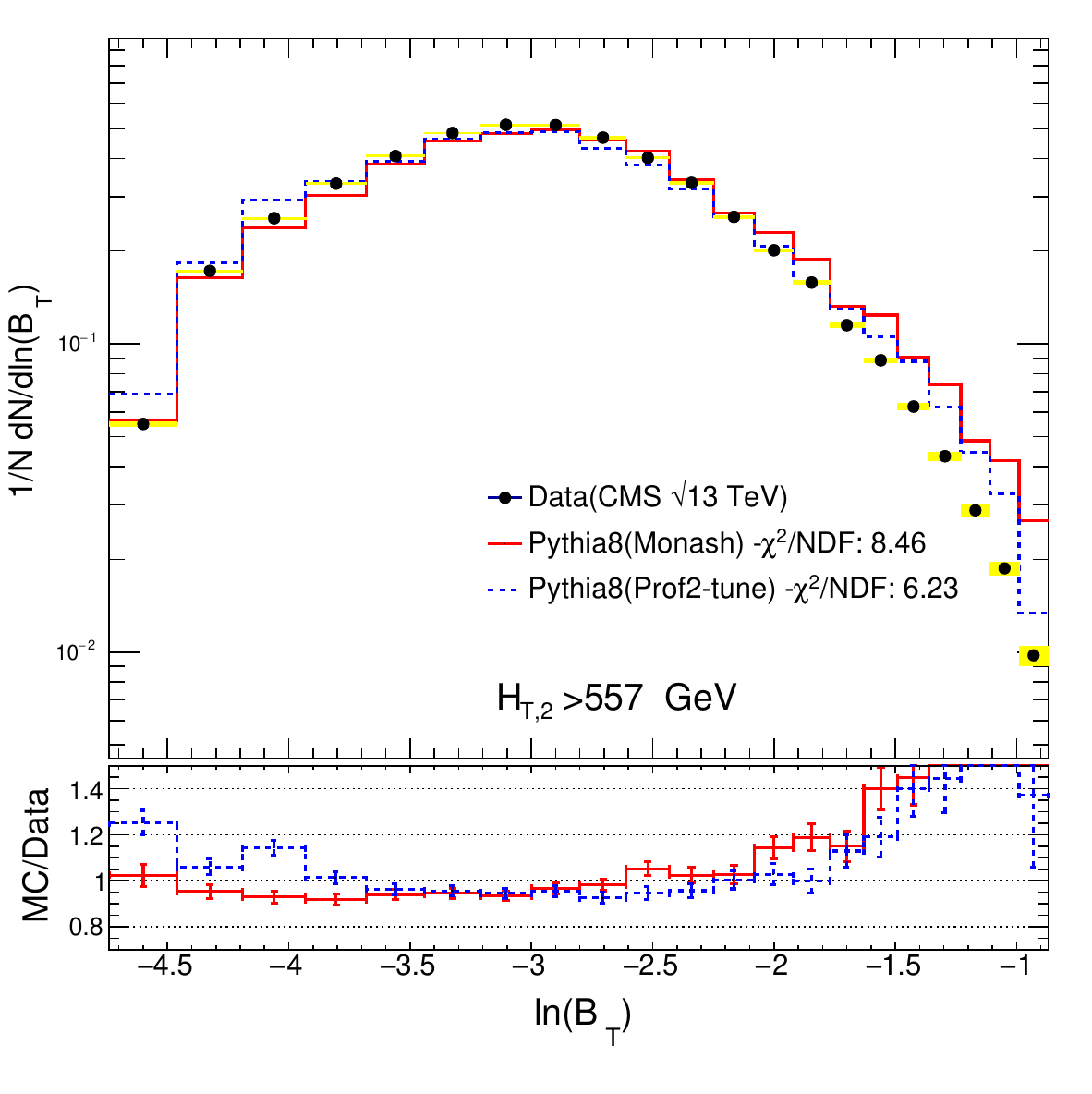}
\caption{\PY with the optimized parameter set is compared with CMS data. Monash tune of \PY 
is also shown for comparison. The plots show normalized distributions of the total jet broadening 
($\mathrm{\bT}$) for eight $\HT$ ranges and the bottom panel in each plot shows 
ratios of MC with data.}
\label{Fig:Monash_broad}
\end{figure}
\vspace{2cm}
\begin{figure}[tbp]
\centering
  \includegraphics[page=25, width=.38\textwidth, height =.37\textwidth ]{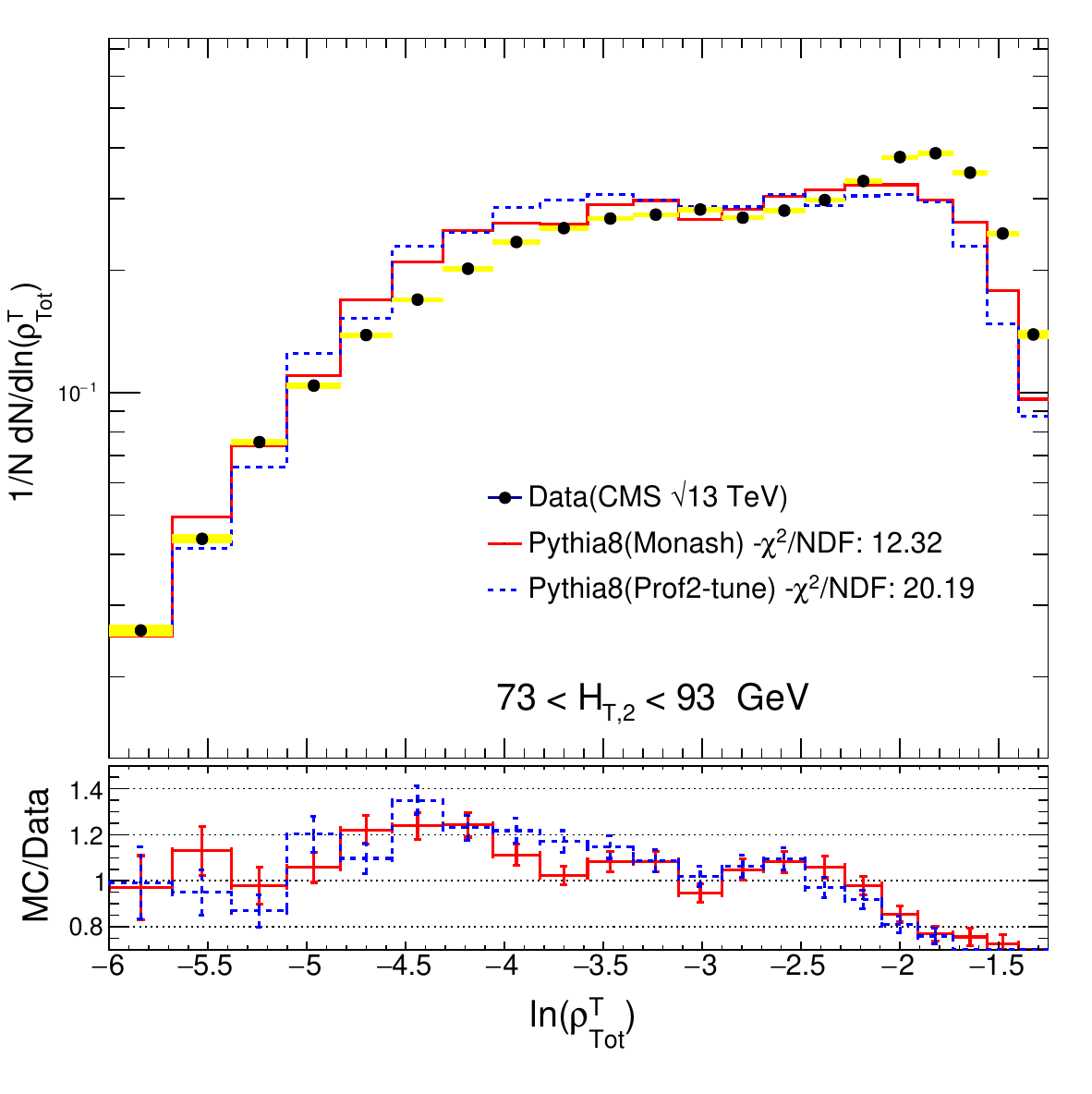}
  \hspace{.5cm}
  \includegraphics[page=26, width=.38\textwidth, height =.37\textwidth ]{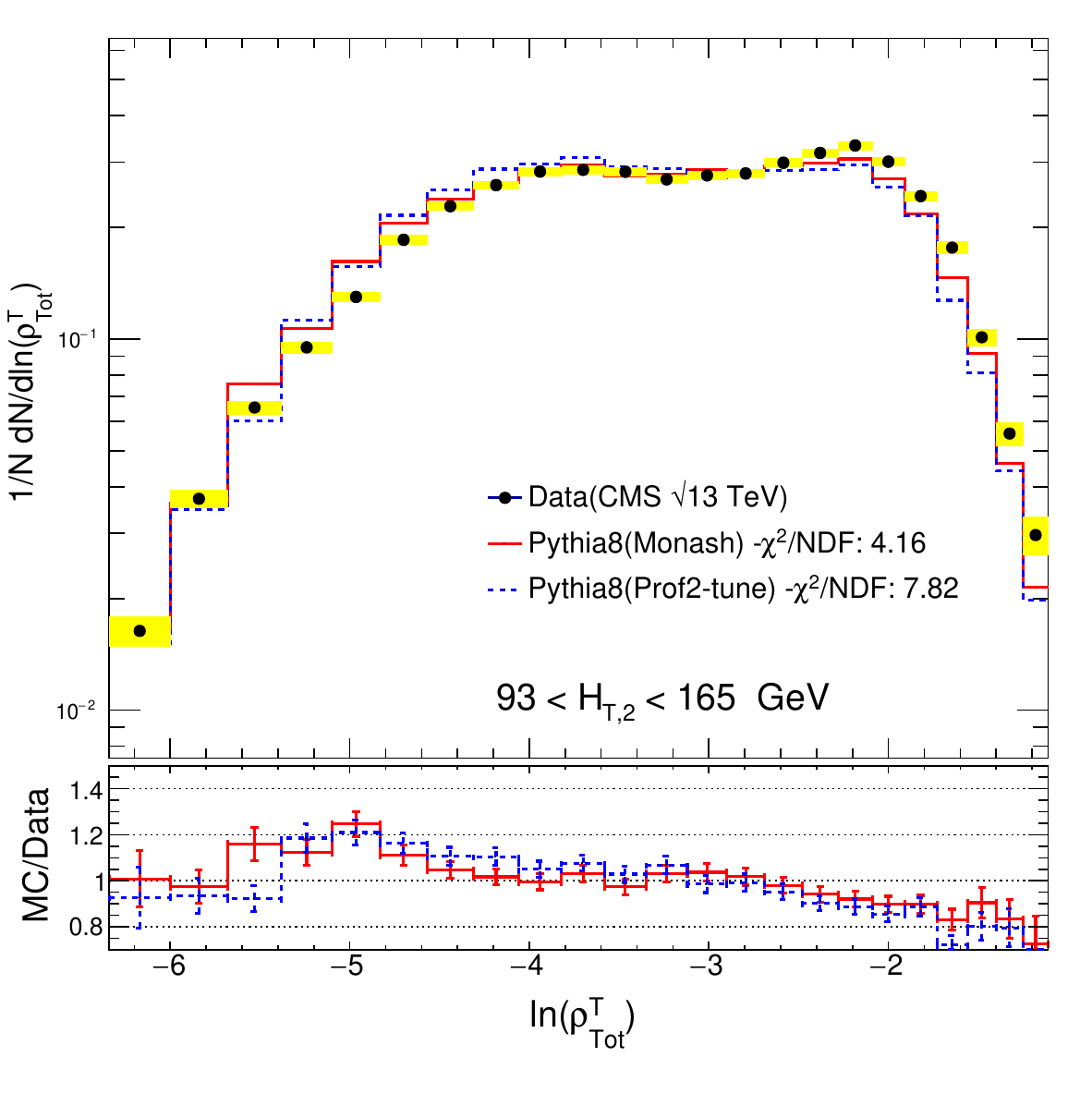}
  \includegraphics[page=27, width=.38\textwidth, height =.37\textwidth ]{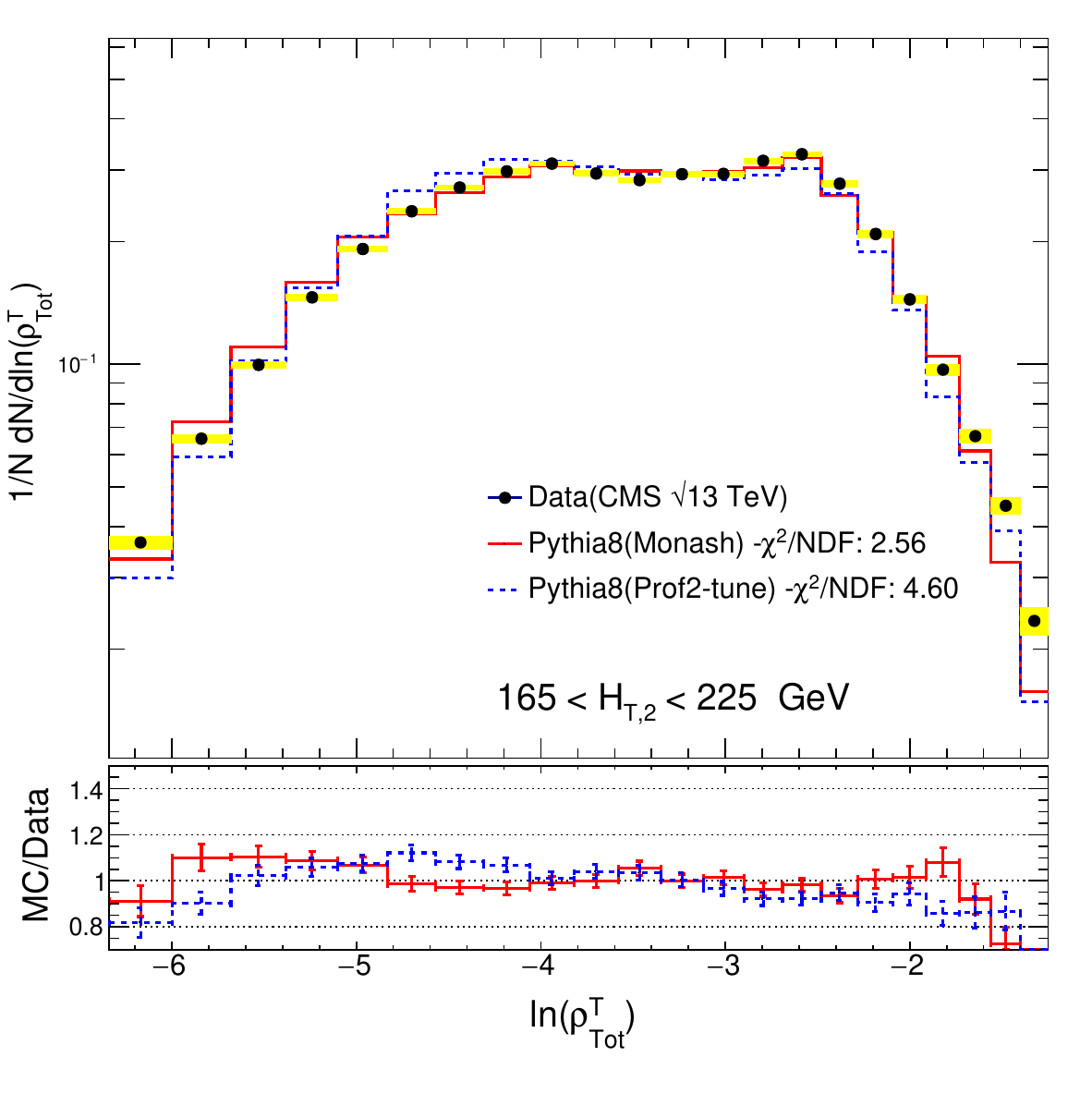}
  \hspace{.5cm}
  \includegraphics[page=28, width=.38\textwidth, height =.37\textwidth ]{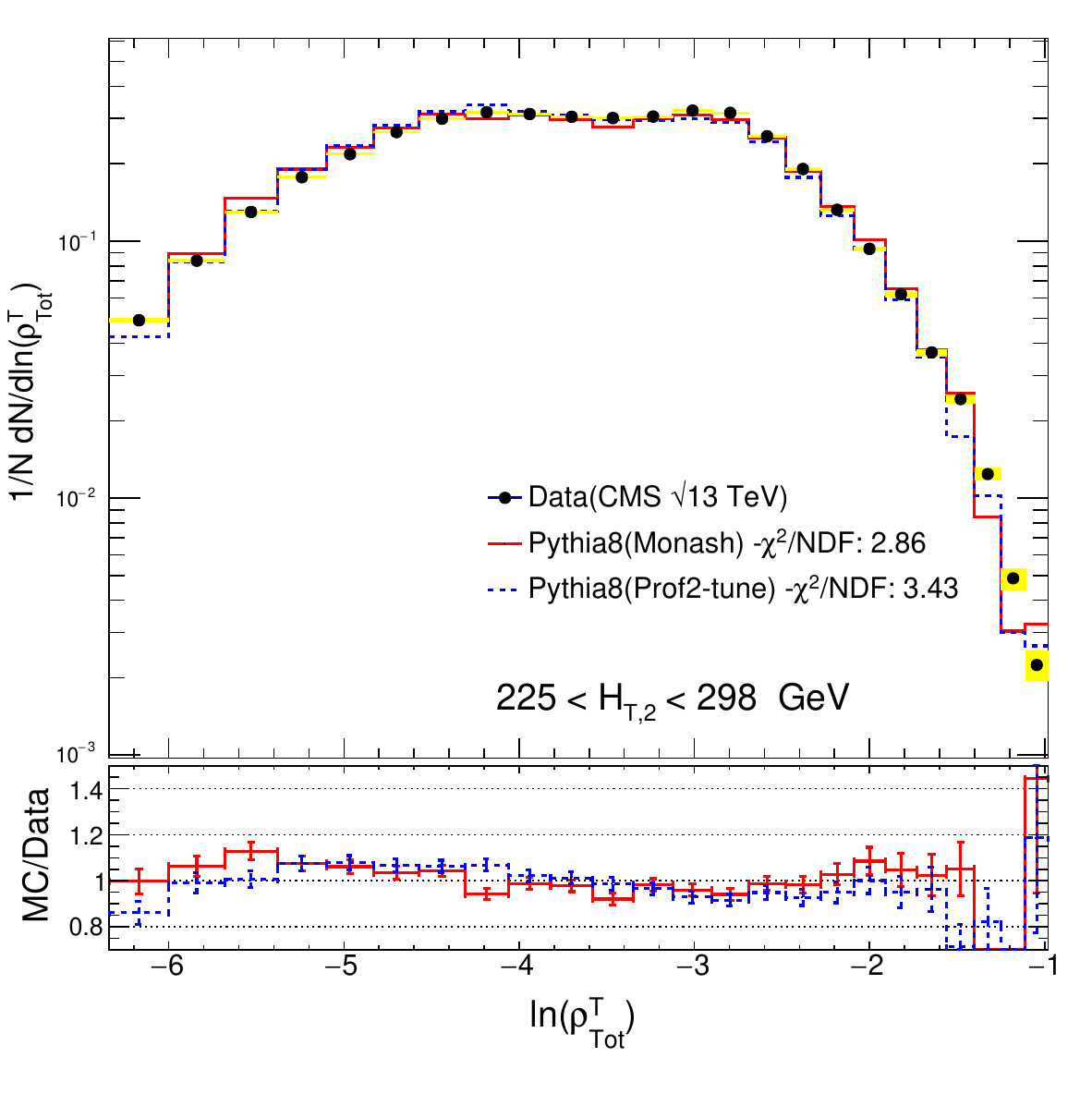}
  \includegraphics[page=29, width=.38\textwidth, height =.37\textwidth ]{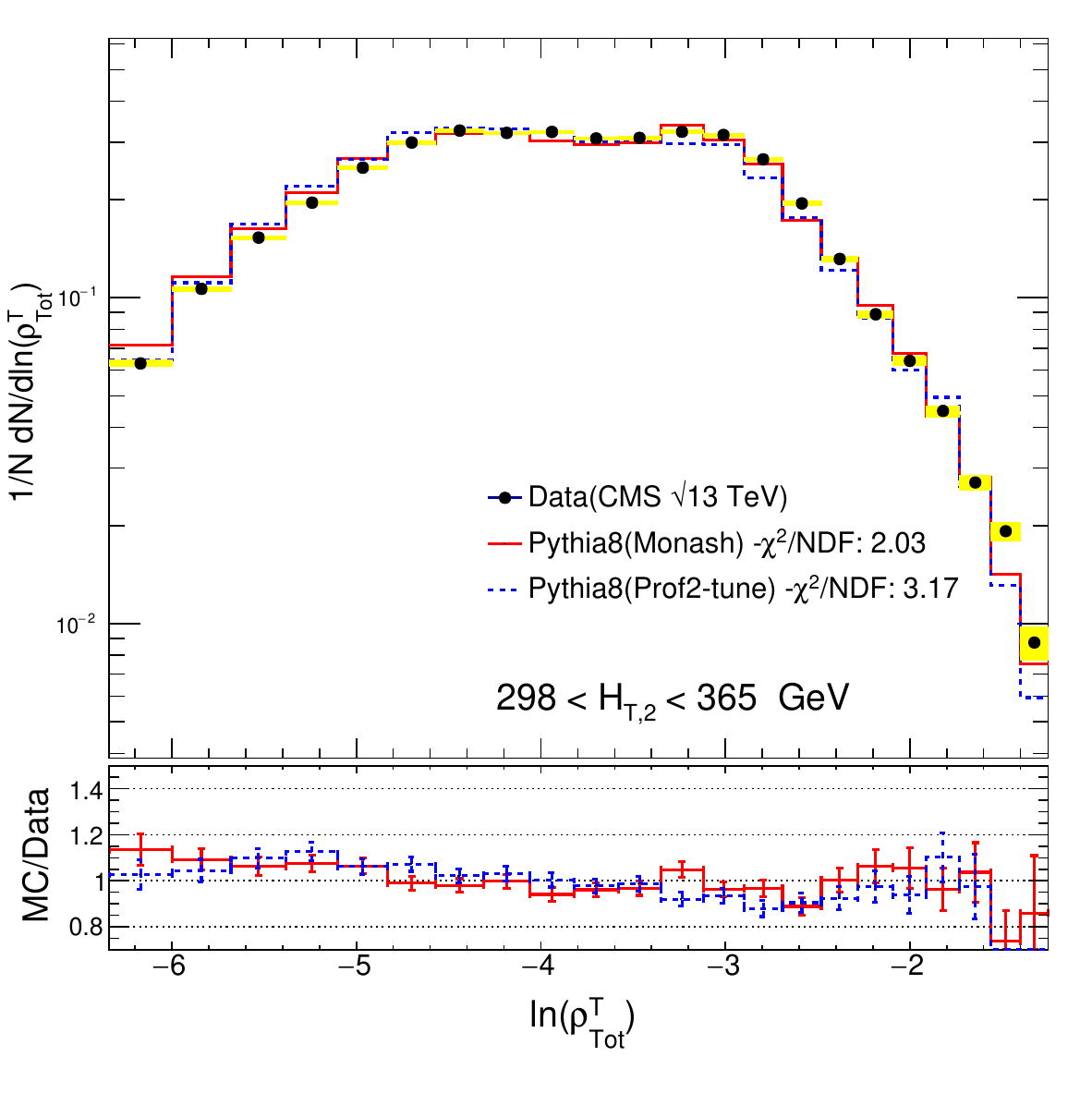}
  \hspace{.5cm}
  \includegraphics[page=30, width=.38\textwidth, height =.37\textwidth ]{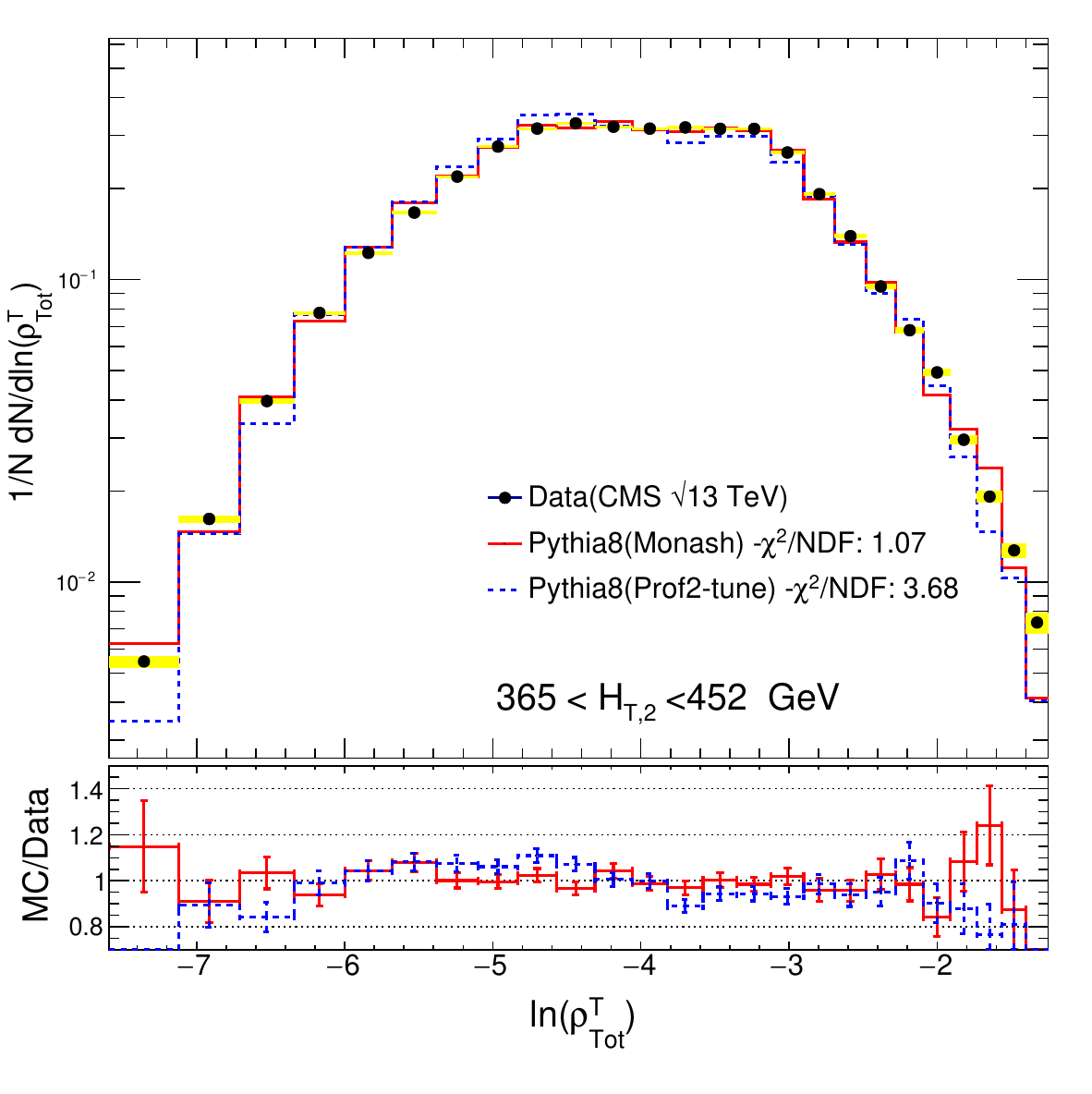}
  \includegraphics[page=31, width=.38\textwidth, height =.37\textwidth ]{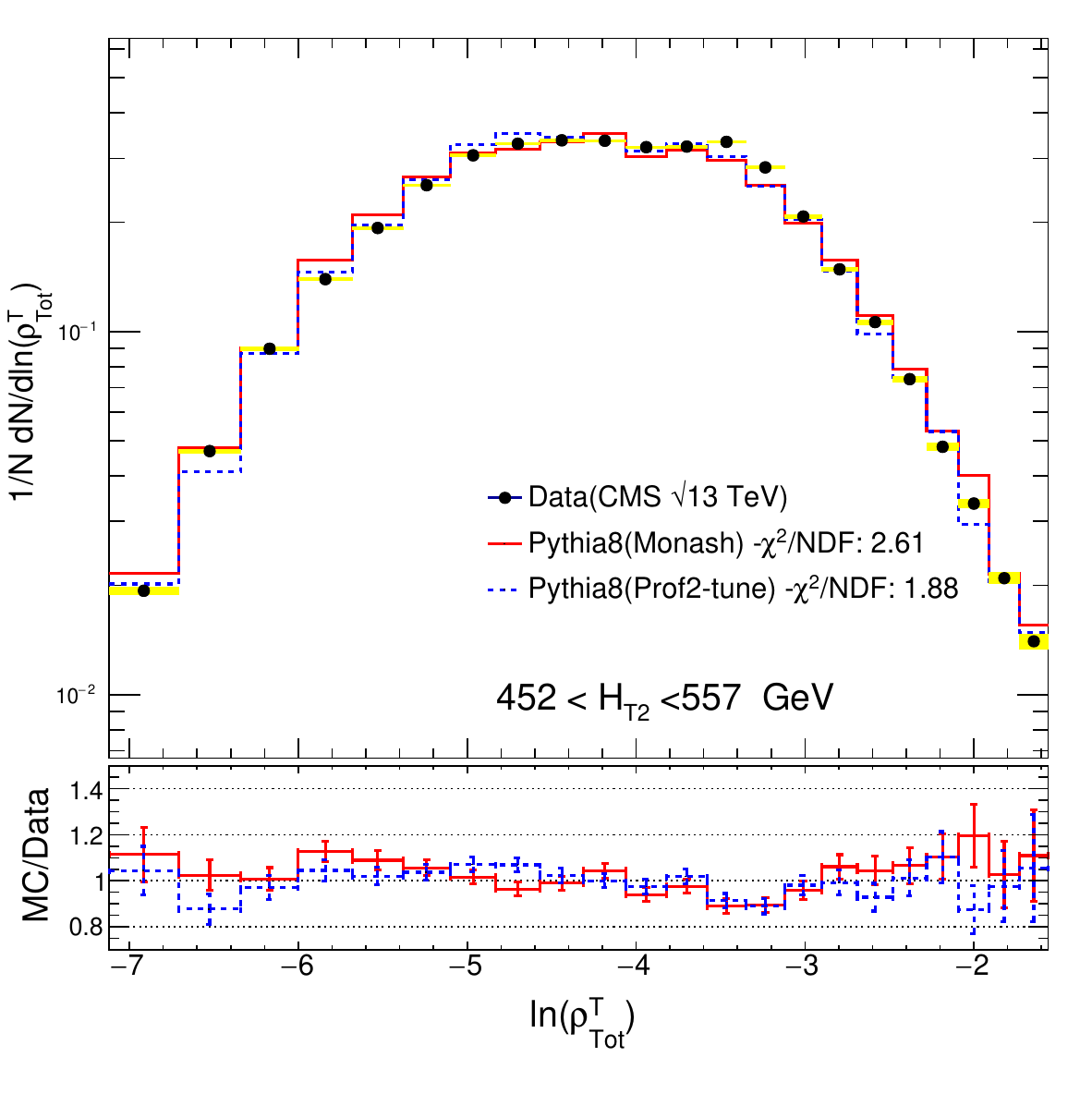}
  \hspace{.5cm}
  \includegraphics[page=32, width=.38\textwidth, height =.37\textwidth ]{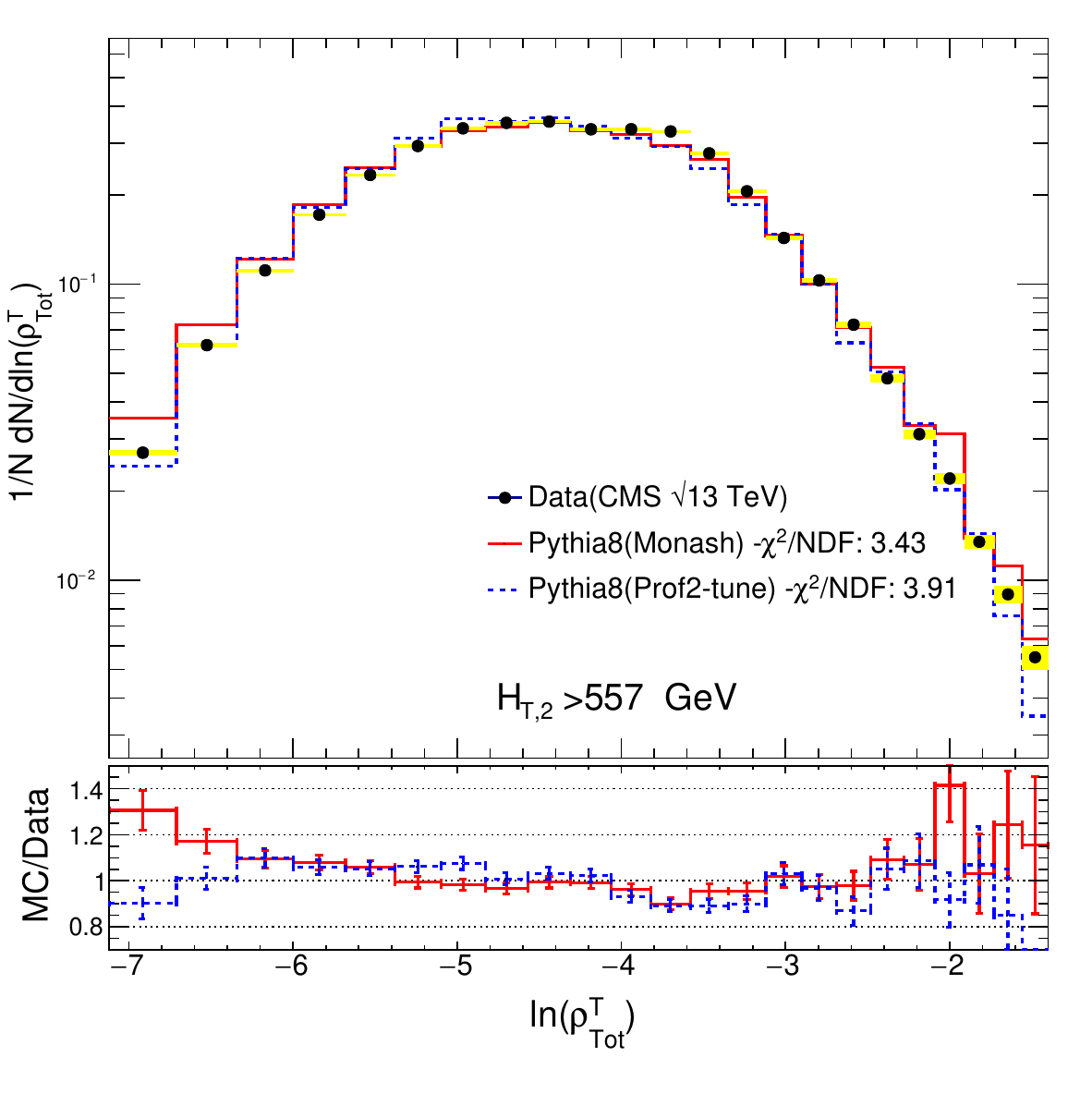}
  \caption{\PY with the optimized parameter set is compared with CMS data. Monash tune of \PY 
is also shown for comparison. The plots show normalized distributions of total transverse jet 
mass($\mathrm{\massp}$) for eight $\HT$ ranges and the bottom panel in each plot shows 
ratios of MC with data.}
\label{Fig:Monash_massp}
\end{figure}
\subsection{Validation of the Optimized Set of Parameters of PYTHIA8} 
The optimized values of the three parameters (see, table \ref{tab:Param1}) are used to calculate the ESVs. 
It is seen that the agreement with data deteriorates slightly for $\mathrm{\taup}$ (figure 
\ref{Fig:Monash_taup}) and $\mathrm{\massp}$ (figure \ref{Fig:Monash_massp}) compared to the 
good agreement with the Monash tune. But, there is significant improvement in agreement with data 
for $\mathrm\mass$ (figure \ref{Fig:Monash_mass}) and $\mathrm{\bT}$ (figure \ref{Fig:Monash_broad}) 
compared to the Monash tune. Since $\mathrm\mass$ and $\mathrm{\bT}$ had a rather poor agreement 
between data and the Monash tune, overall this new set of parameters is better. 

\begin{figure}[tbp]
        \centering
\includegraphics[page=8, width=.38\textwidth, height =.354\textwidth ]{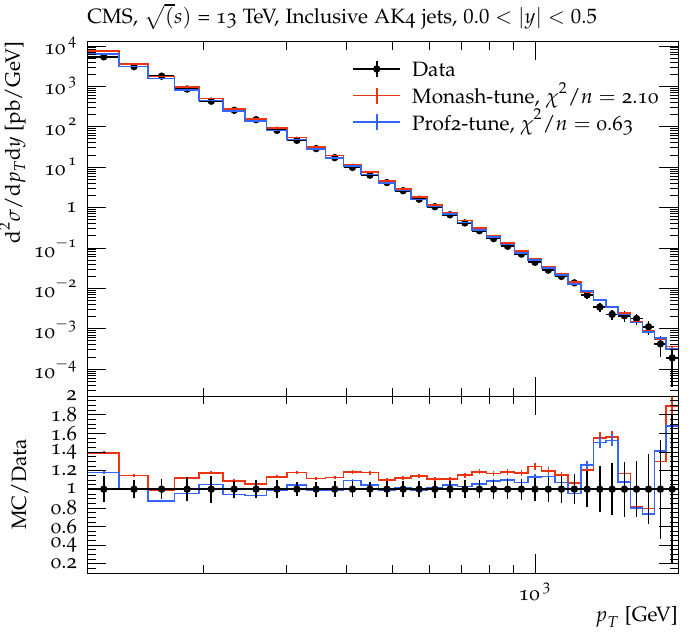}
\hspace{0.4cm}
\includegraphics[page=9, width=.38\textwidth, height =.354\textwidth ]{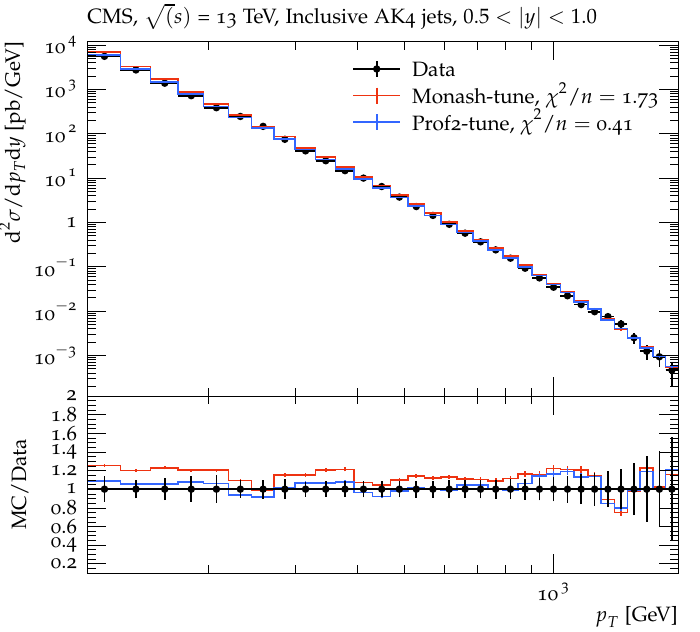}
\\
\includegraphics[page=10, width=.38\textwidth, height =.354\textwidth ]{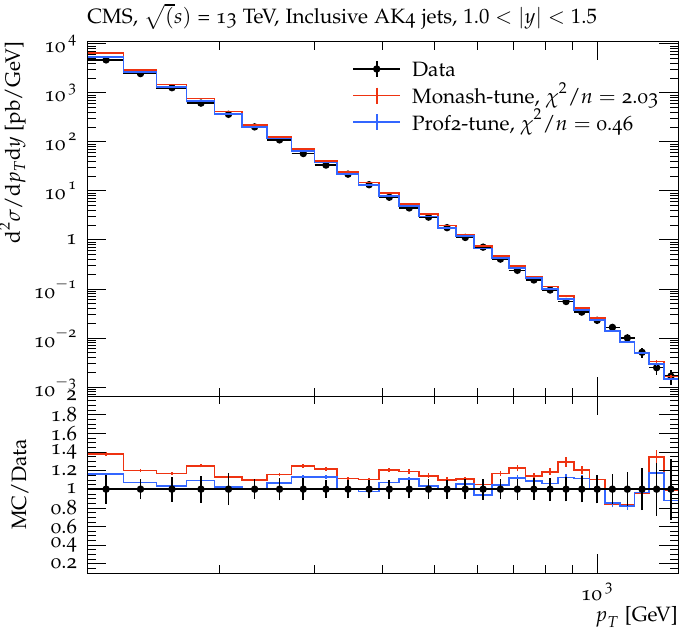}
\hspace{0.4cm}
\includegraphics[page=11, width=.38\textwidth, height =.354\textwidth ]{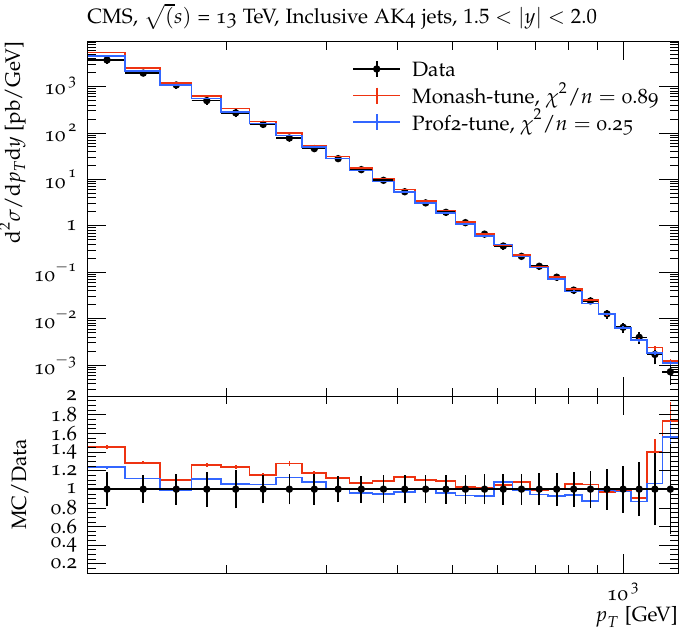}
\\
\includegraphics[page=12, width=.38\textwidth, height =.354\textwidth ]{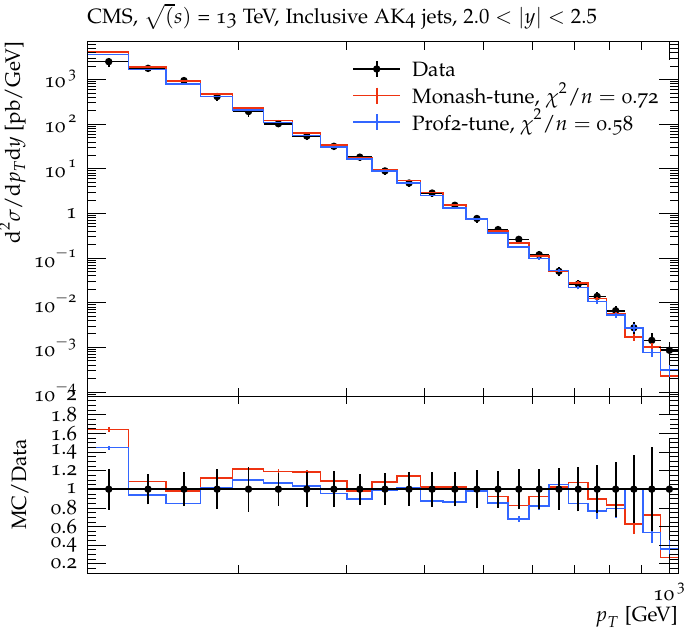}
\hspace{0.4cm}
\includegraphics[page=13, width=.38\textwidth, height =.354\textwidth ]{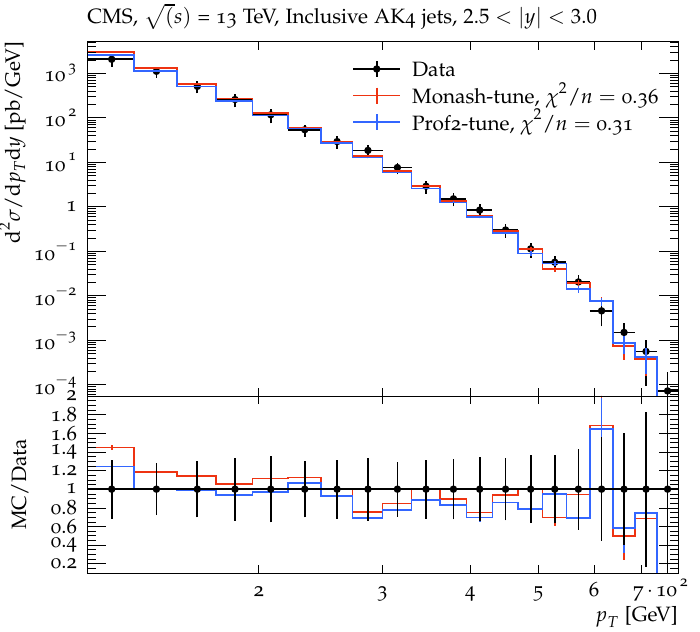}
\\
\includegraphics[page=14, width=.38\textwidth, height =.354\textwidth ]{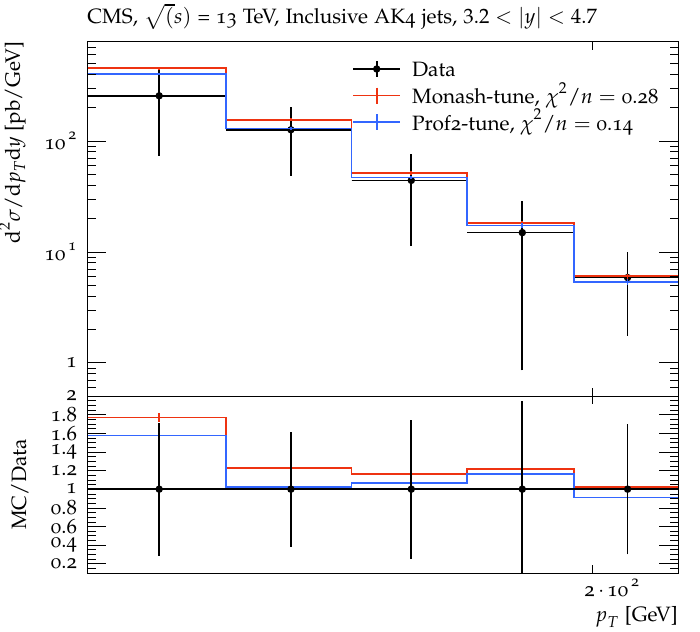}
\caption{\PY with the optimized parameter set is compared with CMS data. Monash tune of \PY 
is also shown for comparison. The plots show normalized distributions of differential inclusive 
cross-section for $\ak$ jets (R=0.4) and the bottom panel in each plot shows ratios of MC with data.}
\label{Fig:CMSinc_jet1}
\end{figure}

\begin{figure}[tbp]
        \centering
\includegraphics[page=1, width=.38\textwidth, height =.354\textwidth ]{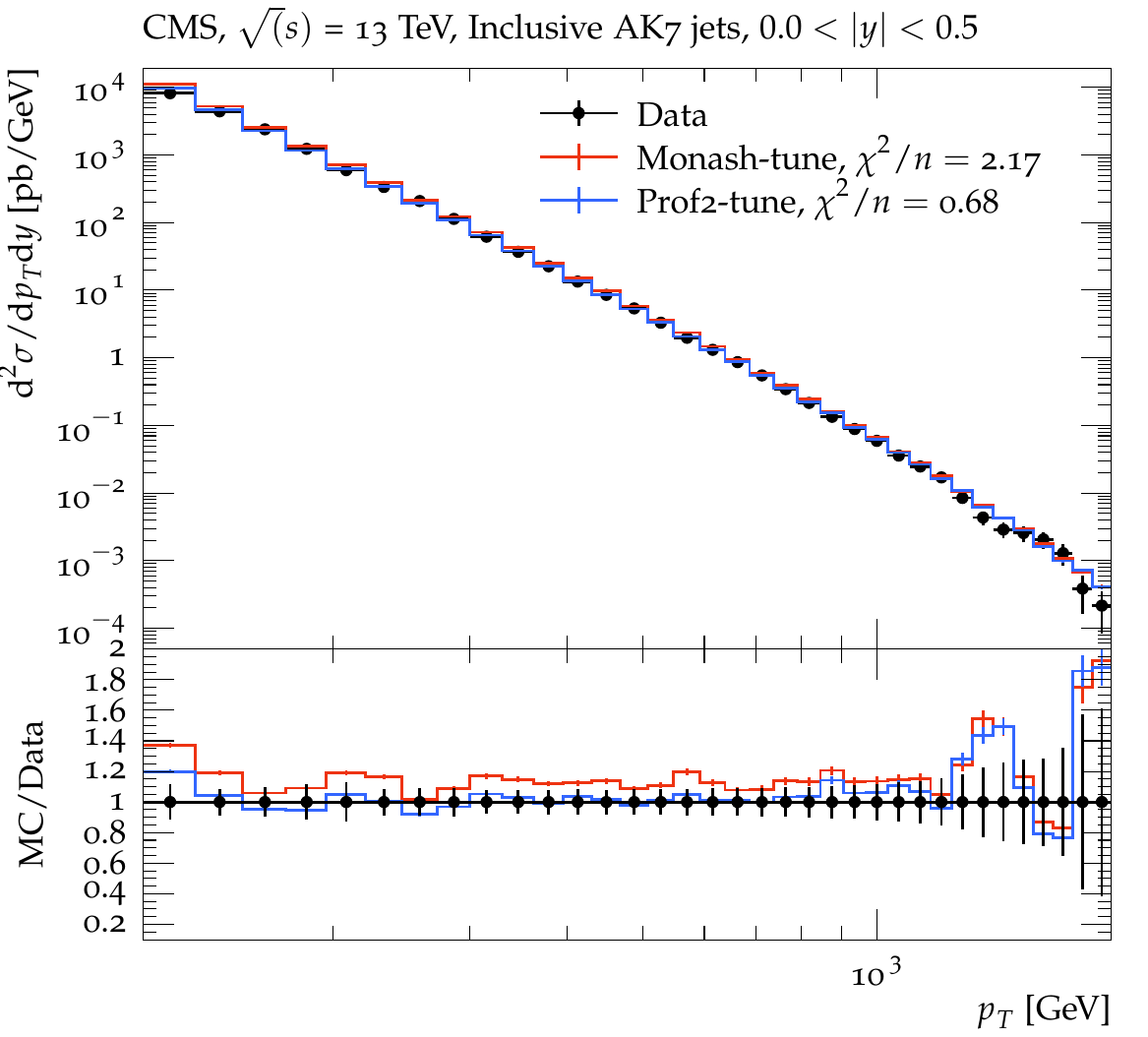}
\hspace{0.4cm}
\includegraphics[page=2, width=.38\textwidth, height =.354\textwidth ]{CMS_2016_I1459051_v3.pdf}
\\
\includegraphics[page=3, width=.38\textwidth, height =.354\textwidth ]{CMS_2016_I1459051_v3.pdf}
\hspace{0.4cm}
\includegraphics[page=4, width=.38\textwidth, height =.354\textwidth ]{CMS_2016_I1459051_v3.pdf}
\\
\includegraphics[page=5, width=.38\textwidth, height =.354\textwidth ]{CMS_2016_I1459051_v3.pdf}
\hspace{0.4cm}
\includegraphics[page=6, width=.38\textwidth, height =.354\textwidth ]{CMS_2016_I1459051_v3.pdf}
\\
\includegraphics[page=7, width=.38\textwidth, height =.354\textwidth ]{CMS_2016_I1459051_v3.pdf}
\caption{\PY with the optimized parameter set is compared with CMS data. Monash tune of \PY 
is also shown for comparison. The plots show normalized distributions of differential inclusive 
cross-section for $\ak$ jets (R=0.7) and the bottom panel in each plot shows ratios of MC with data.}
\label{Fig:CMSinc_jet2}
\end{figure}

\begin{figure}[tbp]
\centering
\includegraphics[page=1, width=.42\textwidth ]{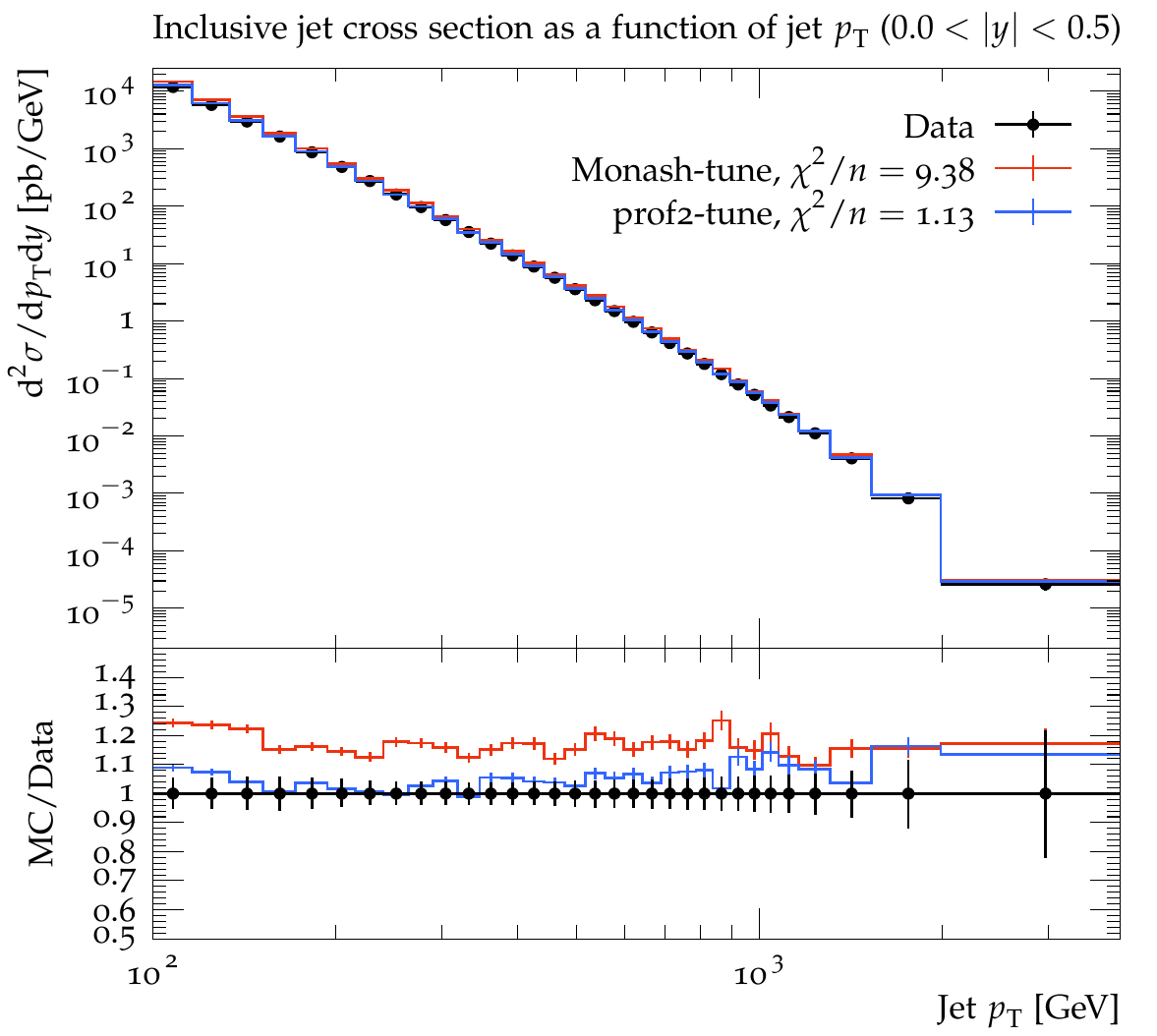}
\hspace{0.2cm}
\includegraphics[page=2, width=.42\textwidth]{ATLAS_2016_CONF_2016_092_V3.pdf}
\includegraphics[page=3, width=.42\textwidth ]{ATLAS_2016_CONF_2016_092_V3.pdf}
\hspace{0.2cm}
\includegraphics[page=4, width=.42\textwidth ]{ATLAS_2016_CONF_2016_092_V3.pdf}
\includegraphics[page=5, width=.42\textwidth ]{ATLAS_2016_CONF_2016_092_V3.pdf}
\hspace{0.2cm}
\includegraphics[page=6, width=.42\textwidth  ]{ATLAS_2016_CONF_2016_092_V3.pdf}
\caption{\PY with the optimized parameter set is compared with ATLAS data. Monash tune of \PY 
is also shown for comparison. The plots show normalized distributions of differential inclusive 
cross-section for $\ak$ jets (R=0.4) and the bottom panel in each plot shows ratios of MC with data.}
\label{Fig:ATLAS_inc_jet1}
\end{figure}

It is imperative that the optimized set of parameters is tested against some other QCD results at 
the LHC. Inclusive jet cross-section is a very important QCD measurement at the LHC and is sensitive 
to PDF of protons and $\mathrm{\alpS}$ and both CMS~\cite{Khachatryan:2016wdh} and 
ATLAS~\cite{ATLAS:2016djc} have studied this with the 13 TeV data. The CMS has measurements of 
inclusive cross-sections for $\ak$ jets with R = 0.4, 0.7. Figures \ref{Fig:CMSinc_jet1} and 
\ref{Fig:CMSinc_jet2} show that the new parameter set improves the agreement with data compared 
to the Monash tune. Similar improvement is seen for the ATLAS measurement of $\ak$ jets with 
R = 0.4 (figure \ref{Fig:ATLAS_inc_jet1}).

\begin{table}[tbp]
    \centering
    \setlength\tabcolsep{4pt}
    \begin{tabular}{|c|cc|cc|cc|cc|}
\hline
$\HT$ range & \multicolumn{2}{|c|}{$\mathrm{\taup}$} & \multicolumn{2}{|c|}{$\mathrm{\mass}$} &
                                        \multicolumn{2}{|c|}{ $\mathrm{\bT}$} & \multicolumn{2}{|c|}{$\mathrm{\massp}$} \\
\cline{2-9} in GeV & Monash & This  & Monash & This & Monash & This & Monash& This \\
              & tune & study &tune & study &tune &study & tune & study \\
            \hline
73-93       & 8.49  & 16.40 & 38.11&16.86 &69.84&22.12& 12.32&20.19\\
93-165      & 3.90  & 9.96 & 19.44 &7.31 &23.79&6.49&4.16&7.82 \\
165- 225    & 1.14  & 5.39 &13.63 &4.38 &21.30&7.17&2.56&4.60 \\
225-298     & 1.81  & 4.59 &11.44 &3.58 &23.51&7.42&2.86&3.43 \\
298-365     & 1.54  & 3.53 &6.29 &1.68 &12.19&3.78&2.03&3.17\\
365-452     & 1.36  & 3.95 &7.42 &2.29 &8.53&5.15&1.07&3.68\\
452-557     & 1.72  & 2.40 &6.29 & 2.09&10.10&4.56&2.61&1.88\\
$>557$      & 1.16  & 3.88 &6.38 & 2.76 &8.46&6.23&3.43&3.91\\

\hline
    \end{tabular}
    \caption{The goodness-of-fit ($\mathrm{\cindf}$) for each ESV corresponding to the Monash tune (left column)
    and the present optimization (right column) are shown for each $\HT$ range.}
    \label{tab:ESVs_ch2}
\end{table}

\begin{table}[htbp]
\centering
    \setlength\tabcolsep{11pt}
\begin{tabular}{|c|cc|cc|cc|}
\hline
    & \multicolumn{2}{|c|}{CMS(R$=0.4$)} & \multicolumn{2}{|c|}{CMS(R$=0.7$)} & \multicolumn{2}{|c|}{ATLAS(R$=0.4$)}\\
        \cline{2-7}  Rapidity & Monash   & This & Monash    &  This & Monash    & This \\
    range  & tune     & study      & tune      & study & tune      & study \\
        \hline
$0.0 < | y | < 0.5$    &    2.10   &   0.63  &    2.17    & 0.68   & 9.38 & 1.13 \\
$0.5 < | y | < 1.0$    &    1.73   &   0.41  &    2.11    & 0.50   & 8.20 & 0.96\\
$1.0 < | y | < 1.5$    &    2.03   &   0.46  &    3.02    & 0.97   & 9.33 & 2.05 \\
$1.5 < | y | < 2.0$    &    0.89   &   0.25  &    1.30    & 0.40   & 6.92  & 1.56\\
$2.0 < | y | < 2.5$    &    0.72   &   0.58  &    1.21    & 0.76   & 5.64  & 1.39 \\
$2.5 < | y | < 3.0$    &    0.36   &   0.31  &    0.27    & 0.15   & 3.75  & 1.23\\
$3.2 < | y | < 4.7$    &    0.28   &   0.14  &    0.28    & 0.23  & - &- \\
\hline
\end{tabular}
\caption{Optimized \PYTHIA shows better agreement in terms of $\mathrm{\cindf}$ values with the CMS measurement of inclusive jet cross
        section for $\ak$ jets with R=0.4, 0.7  and also for ALTAS measurement of inclusive jet cross section with R=0.4 }
\label{tab:CMS2016_1}
\end{table}

\vspace{2cm}

\section{Summary}\label{sec:5}
The strong coupling and maximum shower evolution scale used in the parton shower model of \PY have  
been optimized using the CMS measurement of four event shape variables over a wide range of 
energy scale of the events. \PY with the optimized parameters shows better agreement with 
inclusive jet cross-section measurements by CMS and ATLAS experiments. This study suggests 
that the models of initial and final state radiations in \PY can be improved to better represent 
various Quantum Chromodynamics related studies in the context of the Large Hadron Collider.\\

\section*{Acknowledgments}
The authors sincerely thank Prof. Peter Z. Skands (Monash University) for useful discussions and suggestions.
Financial support by the Department of Science \& Technology, Government of India, through research grants,
SR/MF/PS-02/2014-VB(G) and SR/MF/PS-03/2014-VB(G) is thankfully acknowledged.



\bibliographystyle{ws-ijmpa}
\bibliography{SMP-17-003}

\end{document}